\documentclass[aps,prd,twocolumn,groupedaddress,nofootinbib,showkeys]%
{revtex4-1}

\usepackage{comment}

\usepackage[T1]{fontenc}
\usepackage{enumerate}
\usepackage{bm}
\usepackage{amsmath}
\usepackage{amssymb}
\usepackage{graphicx}
\newcommand{\R}{\Bbb R}

\newcommand{\scrif}{{\mathcal{I}^{+}}}
\newcommand{\scrip}{{\mathcal{I}^{-}}}

\newcommand{\const}{\mathrm{const}}
\newcommand{\sel}{{\sigma _{\rm el}}}
\newcommand{\smag}{{\sigma _{\rm mag}}}
\newcommand{\z}{{z}}

\DeclareMathOperator{\csch}{csch}
\DeclareMathOperator{\sgn}{sgn}

\begin{document}

\title{Magnetic effects in general relativity:  
quadrupoles, shear and memory}

\author{Adam D. Helfer}

\email[]{helfera@missouri.edu}
\affiliation{Department of Mathematics and Department of Physics \& Astronomy,
University of Missouri,
Columbia, MO 65211, U.S.A.}

\date{\today}

\begin{abstract}
Gravitational effect of ``magnetic type'' --- those having a curl-like character over large spheres --- are investigated, for isolated systems.  The Bondi--Sachs--Newman--Penrose formalism clarifies a number of points, especially related to radiation memory.  It is shown that the ``memory tensor'' is equivalent to the change in Bondi shear, from before to after the emission of radiation.  This means that if magnetic radiation memory is present, at least one of the intervals bracketing the radiation must have non-zero magnetic shear but vanishing radiation.  Such intervals, called here CPMS regimes, are shown to be necessarily non-stationary, however, raising a variety of technical and interpretative issues.  
In linearized general relativity, the gravitational fields due to point magnetic quadrupoles with arbitrary time-dependence are computed, and some of their physics studied.  In the far zone, there is a red-shift effect which could be searched for astrophysically:  light coming from behind a source generating magnetic shear would be red-shifted by an amount varying with the angle around the source of shear, and in the far-field limit this red-shift goes inversely with the impact parameter.
Induction-zone effects are also considered.
An induction-zone memory effect should exist which could possibly be within the reach of laboratory experiments, but no good candidates for astrophysically detectable effects are found.  Also a quadrupole will induce test particles to move in such a fashion as to create an opposing quadrupole, an effect reminiscent of Lenz's law.  
\end{abstract}


\maketitle

\section{Introduction}

Effects of ``magnetic type'' --- that is, those coded in fields which  over large spheres have a curl-like, or ``unnatural parity,'' character ---
form an intriguing and relatively unexplored facet of general relativity.
Probably best known are the possible cosmological gravitational-wave $B$-modes.  But there is another sort of magnetic feature, which should appear
in the asymptotic geometry of 
isolated systems, one associated with {\em Bondi shear}.

The Bondi shear $\sigma$ is a sort of potential for gravitational radiation from an isolated system.  It is defined on future null infinity $\scrif\cong \R\times S^2$, with the sphere being the asymptotic outgoing null directions and the $\R$ factor Bondi retarded time $u$.  
I will write 
an overdot for $\partial _u$.  Gravitational radiation is signaled by $\dot\sigma\not=0$.

The Bondi shear is 
a spin-weight two quantity, and in general it has 
components $\sel$ of electric (or gradient) and $\smag$ of magnetic (or curl) types.\footnote{From now on, these will simply be called the electric and magnetic parts of the shear.  There is only a formal connection with electromagnetism.}  It is also affected by 
gauge transformations:  under a supertranslanslation, a $u$-independent term is added to $\sel$.  
Indeed, in a regime where there is no radiation,
the electric part $\sel$ can be removed by a gauge change.
However $\smag$ is gauge-invariant
(and can be thought of as a purely general-relativistic contribution to the specific [per unit mass] spin of the system \cite{ADH2007}).

This paper will be concerned with magnetic Bondi shear, and also with related induction-zone effects.  It has two aims:  to clarify some of the conceptual issues involved, and to assess the potential for the effects to be measurable in the laboratory or detectable astrophysically.  Before outlining the main results, it will be helpful to give a preliminary discussion of the space--times under consideration.

\subsection{Space--times with magnetic shear}

It should be said at the outset that space--times with $\smag\not=0$ would be significantly asymmetric (since $\smag$ has spin-weight two), and we do not have explicit non-singular examples of them in full general relativity.  
We also do not know of any realistic examples of matter which would plausibly generate large persistent magnetic shears.\footnote{As noted above, the quantity $\smag$ can be regarded as a specific angular momentum, like the Kerr parameter $a$, in which case one naturally compares it to the mass of the system (converted to a length).  It is in this sense which realistic estimates for known systems are small.}  
On the other hand, one {\em would} expect the presence of magnetic shear to be generic.
In short, we do expect magnetic shear to be present in realistic systems, but we have no reason at present to think it would be a large effect, and we do not have good examples in full general relativity.

Within the class of space--times with magnetic shear, there is a distinguished set.  These are the space--times which, at least for some interval of retarded time, have no radiation but nevertheless magnetic shear, that is $\dot\sigma =0$ but $\smag\not=0$.  In such an interval, one could by a supertranslation set $\sel =0$, and then one would have a regime with {\em constant} (in $u$) {\em purely magnetic shear}.
I will call these {\em CPMS regimes} (regardless of the gauge chosen for $\sel$).

Although space--times with CPMS regimes are certainly not generic within the class of all space--times admitting Bondi--Sachs asymptotics, one would expect them to be generic within the class of those with non-radiating periods.  
This would suggest that, given that a system at some stage relaxes to the point where it is not radiating gravitationally, we should usually expect to find it in a CPMS state.
Should this be the case, it would be very important to study CPMS space--times as a class, since we would expect these to be the ones representing real systems, except during periods of radiation.

However, there are good (but not compelling) reasons to think that the situation is more complicated:  that in realistic systems CPMS regimes may well occur but cannot persist indefinitely.
One line of thought supporting this view comes
from thinking about implications for black holes, and this will be explained now.  (Others will emerge from discussion later in this paper.)

One would expect a general collapsing body to have non-zero multipole moments of both parities of many orders.  If it forms a black hole, the argument above would suggest that the hole relaxes to a CPMS state.  This, however, would violate the No-Hair Conjecture, for Kerr--Newman solutions have vanishing magnetic shear.    So either, at least in the case of black holes, CPMS final states are forbidden, or the No-Hair Conjecture is violated in a serious way.

But what mechanism could forbid CPMS final states?  (And so, what would be wrong with the previous argument that these states were expected?)  
Some indication of this comes from thinking about CPMS perturbations of Kerr (or Kerr--Newman).  These would be zero-frequency deformations, but there are fairly good arguments that such perturbations are not stable~\cite{Teukolsky1972}. 
If this is the case, then a gravitationally collapsing solution close to Kerr might approach a CPMS state for some time, but then ``hiccup'' away its magnetic shear as radiation and finally approach the Kerr state.  

In linearized gravity, there are no known examples of sources generating indefinitely persistent constant magnetic shear, and systems which maintain a constant magnetic shear for a period do so by having separated contributions to the angular momentum, the product of the separation and the individual contributions increasing quadratically.  That the only known means of producing CPMS regimes seems so mannered can be taken to suggest that they cannot be maintained indefinitely --- although this is certainly not a compelling argument.

If this principle {\em is} true generally, then 
CPMS states necessarily hold {\em latent} radiative degrees of freedom, which 
will be eventually driven active.  This would mean that all gravitational systems would eventually emit their magnetic shear in magnetic gravitational radiation.  We should expect a population of $B$-modes from this.

The reader may have noticed that I have avoided saying that CPMS regimes might be quasistationary.  We will see later that they {\em cannot} be stationary, and indeed certain of their Newman--Penrose asymptotic curvature coefficients must grow polynomially with $u$.
(This growth is associated with restrictions on where the Bondi--Sachs asymptotics will be accurate; it does not obviously indicate any unphysical divergence or breakdown of the theory.)

This discussion of the possible instability of CPMS regimes is just a sketch of what seem the likeliest alternatives at present.  The real point is that
there is a great deal about the dynamics of solutions with magnetic shear which will ultimately need to be resolved.

One further point of terminology:  It is common to use the terms ``far zone'' and ``radiation zone'' interchangeably.  Here, though, it is best to distinguish them, for the CPMS regimes have no radiation but nevertheless distinctive asymptotic structure.  I will use {\em far zone} to refer to the region in space--time (if it exists) in which the physical geometry is well-modeled by the leading terms in the Bondi--Sachs asymptotic expansions.  If there is radiation, the leading term will be the radiation field,  and in the far zone this will appear as outward-directed transverse waves.
But if no radiation is present, then we will see that one part of the leading term is directly due to $\smag$, and uncovering physical consequences of this will be an important part of this paper.\footnote{One might think that, if no radiation is present, the definition adopted here would make the far zone the same as the induction zone.  However, while there is some blurring of the concepts, they are not the same, because many of the asymptotic observables of interest are nonlocal.  An important example will be scattering, which brings in far-zone effects one would not normally consider inductive.}

\subsection{Main results}

I will 
present here the results which most directly bear on questions which have been raised elsewhere or seem most accessible to experiment or observation.
The subsection after this one outlines the paper's contents, including some further results.

\subsubsection{Asymptotic structure and radiation memory.}

Section~II shows that the Bondi--Sachs--Newman--Penrose asymptotic formalism considerably clarifies the structures arising in gravitational radiation memory. 
(A similar perspective, although not going quite as far, is found in the recent paper of Bieri and Garfinkle~\cite{BieriGarfinkle2014}.)  In particular, the ``memory tensor'' is shown to be equivalent to the change in Bondi shear, from before the emission of radiation to afterwards.

An important point brought out by this 
analysis is that, while CPMS regimes by definition have no gravitational radiation, {\em CPMS regimes cannot be stationary}.  In fact, the asymptotic Newman--Penrose equations show directly that certain of the
coefficients in the asymptotic expansion of the curvature
(the Newman--Penrose $\Psi _1^0$, $\Psi _0^0$) {\em grow polynomially} in retarded time $u$ (in a Bondi--Sachs frame).  

It is important to make clear that this does not 
establish that the curvature itself grows unboundedly in any physically meaningful sense, or that the theory breaks down.  
The growth does strongly suggest, however, that {\em as $u$ increases one must go to larger and larger values of the ``distance'' coordinate $r$ to remain in far zone, where the field is well-represented by the leading term in the Bondi--Sachs expansion}.  We may say briefly that {\em the far zone is itself not stationary}, but tends to recede as $u$ increases.\footnote{In fact, the polynomial growth, and the failure of the far zone to be stationary, may occur even if $\smag =0$, but it is unavoidable if $\smag\not= 0$.}

This has a number of implications for radiation memory, especially magnetic radiation memory:

(a) By definition, radiation memory effects involve comparing two non-radiative periods separated by a radiative one.  For magnetic effects, these non-radiative periods must be CPMS regimes (or one of them could have $\sigma$ pure gauge).\footnote{There is a potential issue of terminology here, in that some people might prefer to use ``memory effects'' only for transitions between stationary states; then the result that CPMS regimes cannot be stationary would rule out magnetic radiation memory.  However, this definition would also rule out many interesting cases of the conventional (electric) kind.}  

(b) Magnetic radiation memory effects should be possible, at least over finite periods of retarded time, although if CPMS regimes are unstable they will not persist indefinitely.

(c) M\"adler and Winicour \cite{MaedlerWinicour2016} argued that physically reasonable sources in linearized gravity would not generate magnetic radiation memory.  However, their argument depends on looking at (and making assumptions about) what happens in passing from $u=-\infty$ to $u=+\infty$.  From the present perspective, those results do not speak to what might occur over finite times (and M\"adler and Winicour make stronger assumptions than we would like about infinite times).
(Closely related criticisms were made by Satishchandran and Wald~\cite{SW2019a}.)

(d) The non-stationarity of the far zone means that care is needed in devising observations which could detect magnetic memory.  (For example, one needs to be concerned about whether the far zone, during the course of some proposed measurement, will recede beyond the measuring apparatus.)  In a simple example, we will see that the usual test-mass approaches are problematic, and that certain red-shift measurements would
be a better way to probe magnetic memory.

\subsubsection{A possible laboratory effect.}
The case of a pure magnetic point quadrupole in linearized gravity will be studied in some detail.    It turns out that there is a sort of induction-zone memory effect:  particles near the source will acquire a displacement proportional to the time-integral of the quadrupole.  

That induction-zone memory effects exist is not surprising.  What is interesting is that it might be possible to detect a magnetic one on laboratory scales. One could create a mechanical quadrupole by two parallel oppositely-spinning hoops, and position this near one test-mass of a laser interferometer, and then attempt to measure the growth in the displacement of the test mass.  
Doing this would require adequate mechanical isolation of the quadrupole from the interferometer as well as stability of the interferometer over the run time (or, more precisely, strategies for taking these issues into account insofar as measurements of the displacement go).

For a quadrupole created by two $10^6\, {\rm g}$ hoops of radii $3\times 10^2\, {\rm cm}$, separated by $10^2\, {\rm cm}$ and spinning with angular frequencies $10^3\, {\rm s}^{-1}$, at a distance $10^4\, {\rm cm}$ from the test mass, I estimate displacements of order $10^{-13}\, {\rm cm}\, {\rm y}^{-1}$.

\subsubsection{A possible astrophysical redshift effect.}
The Newman--Penrose asymptotic formulas imply that a CPMS regime contributes to the far-field curvature at the same order as does a mass monopole.  This suggests looking for magnetic effects which are of the same order as the far-field deflection of light by a mass.  
We will see here that (in the case of quadrupole sources in linearized gravity) such an effect does exist, a red-shift.  In fact, a red-shift contribution from $\smag$ exists at this order, whether ${\dot\sigma}_{\rm mag}$ vanishes or not.  (The CPMS condition need not hold.)

The red-shift will apply to light (from distant sources) which passes the quadrupole on its way to the observer.
In the simplest case,
in the limit of large impact parameter $b$,
it will go like $|\smag (u_0)|(\cos\alpha )/b$, where $u_0$ is the retarded time of detection of the light-ray,
and 
the angle $\alpha$ determines where in the plane of the sky around the source the ray passes (measured relative to quantities involving the multipole geometry).

A few words about the form of this result are in order.  First, what has been given here is the far-field (large $b$) form.  While in more general circumstances the effect would depend on $\smag (u)$ for $u\leq u_0$, for large $b$ the dominant term depends only on $\smag (u_0)$; this might be called an {\em amnesiac} characteristic.  Second, the overall $1/b$ dependence here should be compared with the $\sim M/b$ angular deflection of light by a mass $M$; ultimately, this is because the two effects are due to the magnetic and electric parts of the same Newman--Penrose curvature quantity ($\Psi _2$).

The simplicity and long-range character of this form make it an attractive candidate for astrophysical searches.  However, three cautionary points should be noted:  First, a boosted monopole source creates an electric dipole term which also gives rise to a red-shift with a sinusoidal angular dependence, so one needs an independent measurement of the source velocity to adjust for this.  
Related to this, we also need an investigation of possible redshifts due to higher electric multipoles, to see whether the magnetic quadrupole effect could really be distinguished on the basis of observational data.  Finally,
we have at present no convincing argument for any significant $\smag\not=0$ for known astrophysical sources, and so no positive reason to expect any effect --- but this means detection of one would signal interesting and possibly new physics.

\subsubsection{Black holes.}
I pointed out above that Kerr--Newman black holes have vanishing magnetic shear, so one could try to test the No-Hair Conjecture by looking for $\smag$ near black holes.  
Even if the conjecture is true,
black holes might temporarily have magnetic shear, as a result of asymmetric processes in their formation or in accretion to them, which is then radiated.

The Event Horizon Telescope collaboration\footnote{https://eventhorizontelescope.org/}  has reported that the near and induction zones of the supermassive black hole in M87 can be modeled by a Kerr geometry with an accretion disc, and in fact the data have been used to constrain possible {\em electric} quadrupole contributions there \cite{EHT1ab}
(following a suggestion of Johannsen and Psaltis \cite{JP2010}).  
The results of the present paper show that {\em magnetic} quadrupole (or higher) contributions would give rise to secular changes in the geometry.  Detailed modeling would be required to know these in the near and induction zones, but one would expect to be able to constrain these observationally.  Magnetic contributions to the solution, but in the far zone, might also be probed by the red-shift effect described above.

\subsection{Outline}

Section II puts radiation memory effects in the context of the Bondi--Sachs--Newman--Penrose asymptotic formalism.  While some elements of this have been noted previously (see e.g. Bieri and Garfinkle \cite{BieriGarfinkle2014}), the connections between memory, shear and the Bondi--Metzner--Sachs group are brought out very clearly in this language.
With the Newman--Penrose asymptotic formulas, we also see directly that CPMS regimes cannot be stationary.

Section III introduces the simplest example of a magnetic source, a pure point quadrupole in linearized gravity.  The metric can be explicitly computed, as can the change to Bondi coordinates.  A mechanical model producing such a field 
is briefly discussed.  The Bondi shear is shown to go as the second time-derivative ${\ddot Q}_{ab}$ of the magnetic quadrupole moment.  While it is not hard to arrange for the CPMS condition to hold for finite intervals, no plausible mechanism is known which would keep ${\ddot Q}_{ab}$ constant indefinitely.

In Section IV, I consider test particles which are initially comoving with the source.  The ones sufficiently far away are the first candidates to come to mind for magnetic memory effects, but it is shown that these particles do not remain in the far zone (on account its non-stationarity).

However, several {\em induction-zone} effects are explored in Section IV.
One is the memory effect described in Subsection I.B.   Another is reminiscent of Lenz's law:  a distribution of particles, initially at rest around the source, will acquire from the field an induced motion, and the induced magnetic quadrupole of the particles {\em opposes} the original one.

One would like to know whether the induction-zone memory effect could be expected to be detected astrophysically.  Although several consequences are considered, there is little reason to feel encouraged.  There are serious difficulties, and no good arguments in favor.

The first problem is, again, that while there are good reasons to think that generically magnetic effects will exist, we have so far no mechanism identified which would produced substantial ones.  So we have no reason to expect a strong driving force.

Even if induction-zone astrophysical memory 
effects do exist, identifying them would not be easy.  One would have to find some structure in the astrophysical system, near the source, whose past behavior could be inferred confidently enough that the difference could reasonably be ascribed to magnetic effects. (For instance, it turns out that if longitudinal filaments were initially present about the source, magnetic induction memory would tend to distort these to S-shapes.)

Section V derives the red-shift effect described in Subsection~I.B.

Section VI takes up some delicate issues of the structure of the far zone of a CPMS quadrupole. 
The underlying concerns are questions of which aspects of the geometry might be practically measurable.

For instance, I noted above that the far zone will itself typically be receding as retarded time $u$ increases.  It could happen that a test particle is in the zone at one point, and 
is even moving away from the sources, yet is overtaken by the zone's trailing edge --- its trajectory does not remain in the far zone.  One would like to know that there is some suitable  
class of test trajectories which {\em do} lie in the far zone, 
and whose scattering {\em does} give clean information about the asymptotic geometry.
It is shown that this is possible, in the case of linearized quadrupole sources, for a class of distant, relativistic, trajectories (but may fail in less restricted circumstances).

The final section is given to discussion.  The main results are reviewed, and 
the issue of distinguishing between electric and magnetic radiation memory effects is considered.  
The main difficulty in doing this is to get information from at least a measurable fraction of the asymptotic directions around a source.

\subsection{Notation, conventions and background}  
Conventions are as in Penrose and Rindler \cite{PR1984,PR1986}.
The metrics have signature $+{}-{}-{}-$. 
The curvatures are defined by $[\nabla _a,\nabla _b] v^d =R_{abc}{}^d v^c$, $R_{ac} =R_{abc}{}^b$, and Einstein's equation is $R_{ab}-(1/2)Rg_{ab}=-8\pi GT_{ab}$, with $G$ Newton's constant and $T_{ab}$ the stress--energy.  (In this last equation, and throughout, the speed of light is taken to be unity.)  The alternating symbol $\epsilon _{txyz}=+1$ in a right-handed orthochronous frame.

Although the Newman--Penrose formalism is used systematically in the next section, a detailed technical understanding of it is not necessary.  A few of the arguments there do require knowing basic properties of the $\eth$ operator \cite{NP1966,PR1984}.
It is conventional, when dealing with a quantity of non-zero spin-weight, to use $j$ (rather than $\ell$) for the multipole index, and this is done here. 

In Section II, except where otherwise stated, the analysis is valid in full general relativity.  However, in succeeding sections the computations are done in linearized gravity, in standard coordinates.
The Minkowskian metric is $\eta _{ab}$ and the perturbed metric is $g_{ab}=\eta _{ab}+h_{ab}$. 
The vector $t^a=\partial _t$.  It is convenient to adopt the `radiation normalization' $l_at^a =1$ for the null tetrad.  Then in Minkowswki space we have $x^a = ut^a +rl^a$.  
We may think of $u$, $r$ and $l^a$ as the coordinates of the point.  (The null vector is not really a coordinate, of course, but we think of it as determining a point on $S^2$ which we could coordinatize by standard means.)

\section{Shear, memory and sources}

I outline here the main ideas relevant to magnetic memory in the asymptotic formalism developed by Bondi, Sachs, Newman and Penrose.

\subsection{Decompositions by frame and parity}

There are two distinct sorts of decompositions into ``electric'' and ``magnetic'' parts in general relativity.  One, which we have already encountered, will be of primary interest here.
However, the second meaning will come up briefly.  

The main case of interest is a decomposition of a {\em spin-weighted function on the sphere} into two parts with certain parity properties.  
(A familiar example is the resolution of electromagnetic polarizations on the celestial sphere to $E$-modes and $B$-modes.)
If $\lambda$ is a function of spin-weight $s\geq 0$ on the sphere, it can be written as $\lambda =\eth ^s\alpha$ for some complex-valued function $\alpha$.  (Here $\eth$ is a certain first-order differential operator, essentially an anti-holomorphic derivative~\cite{PR1984}.)  The {\em electric } and {\em magnetic} parts of $\lambda$ are then $\lambda _{\rm el}=\eth ^s\Re\alpha$ and $\lambda _{\rm mag} =i\eth ^s\Im\alpha$.\footnote{Newman and Penrose refined the electric part of $\lambda$ to be $\eth ^s\Im\alpha$ (that is, omitting the factor of $i$) \cite{NP1966}.}
It is the need to solve the elliptic equation for $\alpha$ which makes this decomposition {\em non-local} (for $s\not=0$).\footnote{The electric and magnetic parts have sometimes been referred to as even and odd parity, but this is misleading.  (Already in the spin-weight zero case, this notion of parity is {\em not} the ordinary one.)   It would be in keeping with language used elsewhere in physics to say the parts have {\em natural} and {\em unnatural} parity, although these terms are not literally accurate, either.}
On the other hand, one can test whether $\lambda$ is purely electric or magnetic by checking whether ${\overline\eth}^s\lambda$ is purely real or imaginary (respectively).  

The terminology comes from Maxwell's electromagnetism in Minkowski space.  There, an oscillating electric or magnetic multipole source will produce a radiation field of electric or magnetic type.

We will also briefly refer to
 {\em local} decomposition of {\em tensors} relative to a choice of timelike vector.  The case which will come up in this paper is the Weyl tensor $C_{abcd}$ and its electric and magnetic parts $E_{ab}$ and $B_{ab}$. 
In order to avoid confusion with the other sense of electric and magnetic, 
I will call these the {\em frame-electric} and {\em frame-magnetic} parts of $C_{abcd}$. 

Finally, to avoid confusion, it may be worth noting that the term {\em gravitomagnetism} is related to but distinct from the ones above, usually referring to effects which can be traced to components other than the time-time one of the metric (with respect to a chosen frame).

\subsection{Bondi--Sachs asymptotics}

Bondi and coworkers \cite{BVM}, followed by Penrose \cite{Penrose1964} and Newman and Penrose \cite{NP1961} gave a framework for treating gravitational radiation.  An isolated general relativistic system admits certain asymptotics which can be conveniently described by adjoining a null hypersurface $\scrif\cong \{ u\in\R\}\times S^2$ at future null infinity.  Here $u$ is a {\em Bondi retarded time parameter}, and $S^2$ is the sphere of asymptotic null directions.

The coordinate $u$ can be extended inwards to the physical space--time by choosing the $u=\const$ hypersurfaces to be null and meet $\scrif$ orthogonally.  They are then ruled by outgoing null geodesics.  Each (sufficiently distant) point in the space--time will lie on a unique such geodesic, and its angular coordinates are those corresponding to the value on $S^2$ at the end-point of the geodesic.  Finally, one
introduces a coordinate $r$ which is an affine parameter on those geodesics.  There is a frame associated with this, and tensor components are expressed in terms of this.
The frames chosen take advantage of the complex structure on the sphere; the tensor components are generally complex, and have spin-weight.  

A basic result is {\em Sachs peeling}, which governs the behavior of the curvature tensor.  The five complex components of the Weyl tensor are $\Psi _n$ ($0\leq n\leq 4$), going as
\begin{eqnarray}
  \Psi _n \sim \Psi _n^0 r^{n-5} +\cdots\, .
\end{eqnarray}
In particular, the {\em radiative} component is $\Psi _4\sim O(1/r)$, and the {\em semi-radiative} one is $\Psi _3\sim O(1/r^2)$; they are linked by a Bianchi identity $\partial _u\Psi _3^0 =\eth\Psi _4^0$.  

While there is a certain universal asymptotic structure common to all the admissible systems, there is also an infinite-dimensional set of motions preserving that structure.  Those motions form the Bondi--Metzner--Sachs (BMS) group; they are generated by Lorentz motions and {\em supertranslations} $u\mapsto {\acute u} = u+\alpha$, where $\alpha$ is an arbitrary smooth real-valued function on $S^2$.  (The {\em translations} are those supertranslations with $\eth ^2\alpha =0$.)

A key quantity is the {\em Bondi shear} $\sigma$, a spin-weight two function on $\scrif$.\footnote{In the Newman--Penrose formalism, it is denoted $\sigma ^0$, but to avoid clutter I drop the superscript.  However, the superscripts on the curvature quantities will be retained.}  Its derivative $\partial _u\sigma$ signals the presence of gravitational radiation.  In fact this derivative is a potential for $\Psi _4^0$ and $\Psi _3^0$, with 
\begin{eqnarray}
  \Psi _4^0 &=&-\partial _u^2{\overline\sigma}\, ,\label{psifour}\\
  \Psi _3^0 &=& -\partial _u\eth{\overline\sigma}\label{psithree}
  \, .
\end{eqnarray}  
The shear is {\em not} BMS-invariant; under a supertranslation it changes to ${\acute\sigma} = \sigma -\eth^2\alpha$.  
Note that this means $\sel$ changes, but not $\smag$.

\subsection{Memory and shear}

Suppose one has two intervals $I_1$ and $I_2$ of retarded time $u$, in each of which the neighborhood of $\scrif$ is very nearly Minkowskian, but in the interim a gravitational wave has passed.  In each of the regimes the shear will be pure gauge:  we will have $\sigma =\eth ^2\alpha _1$ in $I_1$ and $\sigma =\eth ^2\alpha _2$ in $I_2$, with $\alpha _1$ and $\alpha _2$ supertranslations.  In general, we will have $\alpha _2 -\alpha _1$ a supertranslation (and not merely a translation).  This means that {\em even though the two regimes are individually Minkowskian, the evolution from one to another cannot be asymptotically effected by a Poincar\'e motion}.  In particular, the relative relations between test particles' trajectories will {\em not} be preserved by evolution from $I_1$ to $I_2$.  This is an example of a memory effect, following directly from
the work of Bondi, van den Burg and Metzner \cite{BVM},  but discovered from different perspectives and in different contexts by later authors \cite{ZeldovichPolnarev1974,BraginskyGrishchuk1985,BraginskyThorne1987,Christodoulou1991}.

The assumption, in the previous paragraph, that in the nonradiating regimes $I_1$, $I_2$ the space--time was asymptotically so very nearly Minkowskian that their shears were pure gauge was made for conceptual simplicity.  It had the effect of setting $\smag =0$, but this is not an obviously necessary assumption.  In any nonradiating regime, the leading curvature term in the sense of Sachs peeling will be the Newtonian-order $O(r^{-3})$, whether there is magnetic shear or not.  

Suppose we have the test particles following a congruence of timelike geodesics $\gamma$ in a neighborhood of $\scrif$, that congruence tending to evolution along the time axis of a Bondi system.  Then the geodesic deviation equation ${\dot\gamma}^a\nabla _a{\dot\gamma}^c\nabla _c w^d ={\dot\gamma}^a{\dot\gamma}^cR_{abc}{}^d w^b$ for a connecting vector field $w^a$ (with $R_{abc}{}^d$ the Riemann curvature and $\nabla _a$ the covariant derivative) becomes in the asymptotic regime
\begin{eqnarray}\label{agde}
  {\ddot w}^d &\simeq& (\Psi _4^0 m_b{}m^d+\text{conjugate}) r^{-1}w^b
\end{eqnarray} 
with $l^a$, $m^a$, ${\overline m}^a$, $n^a$ a standard null tetrad (and the dots indicating covariant differentiation along the geodesics).\footnote{This is essentially the same as the argument of Bieri and Garfinkle \cite{BieriGarfinkle2014}.
A subtlety is that asymptotically evolution along the time axis will not be geodesic when radiation is present.  However, this gives only a second-order correction to eq. (\ref{agde}).}
Using the formula (\ref{psifour}), we find for the change in connecting vector over the period of gravitational radiation
\begin{eqnarray}\label{memeq}
\Delta w^d&\simeq & -(\Delta\overline\sigma m_b{}m^d+\text{conjugate}) r^{-1}w^b_0\, ,
\end{eqnarray}
where $w^b_0$ is the initial connecting vector and $\Delta \sigma$ is the change in shear.   
The quantity in the parentheses in eq. (\ref{memeq}) is sometimes called the {\em memory tensor}.  In principle, observations of this memory effect for different asymptotic geodesics and initial connecting vectors determine $\Delta\sigma$.\footnote{A distinction between ``linear'' and ``nonlinear'' memory is sometimes made.  This refers to different mechanisms contributing to the change in shear.  The treatment here does not require this distinction.}

There has been some discussion of whether radiative 
{\em magnetic memory} is possible.  This really involves two questions:  whether $\Delta\sigma$ may have a magnetic component;\footnote{In the geodesic deviation equation (\ref{agde}), the quantity in parentheses (times $r^{-1})$ is the asymptotic value of the {\em frame}-electric part of the Weyl tensor.  But that is, in itself, quite irrelevant to the question of magnetic memory in this regime.} and whether, if such a component is possible, among its consequences is what one may reasonably identify as a memory effect.  
I have already indicated that there is no obvious reason to rule out a magnetic contribution to $\Delta\sigma$.  However, 
there are a number of factors which complicate the situation, and we need to understand more of the geometry.

In regimes for which $\dot\sigma =0$, one of
the asymptotic Newman--Penrose equations reduces to
\begin{eqnarray}
    \Psi _2^0 - \overline{\Psi _2^0} = 
        {\overline\eth }^2\sigma - \eth ^2\overline\sigma\, ,
\end{eqnarray}
and this
implies that $\smag$ encodes {\em precisely} the information in the asymptotic curvature component $\Im\Psi _2^0$ --- the magnetic part of $\Psi _2^0$.  {\em In a non-radiating regime, the magnetic shear determines a magnetic contribution to the curvature at the same power of of $r$ as Newtonian terms.}

Perhaps most importantly, CPMS
space--times  {\em cannot} be stationary.  This follows from the Newman--Penrose equations 
\begin{eqnarray}
 \partial _u\Psi _1^0 &=&-\eth\Psi _2^0\label{npo}\\
 \partial _u\Psi _0^0 &=& -\eth \Psi _1^0 +3\sigma\Psi _2^0
   \label{npt}
\end{eqnarray}
for such space--times.\footnote{There are extra terms if material radiation at infinity is allowed, and in principle in special cases these might lead to certain cancellations.  But the point here is that we generically expect $\Psi _1^0$ and $\Psi _0^0$ to be time-dependent.}  
Because we have seen that $\smag\not=0$ implies $\Psi _2^0$ has  non-trivial $j\geq 2$ contributions, the curvature quantities $\Psi _1^0$ and $\Psi _0^0$ must, in the CPMS case, have time-dependent such terms.  (Such nontrivial higher-multipole terms could well be present in cases of purely electric radiation memory as well --- in fact, generically {\em would} be expected to be present.  However, in the electric case it is at least mathematically self-consistent to assume these multipoles vanish.)

Finally, it is worth noting that the quantities $\sel$ and $\smag$ figure importantly but differently in the treatment of general-relativistic angular momentum.  The electric part contributes directly to the general-relativistic analog of the origin-dependent terms, and this is where the issues with supertranslations enter.  On the other hand, one can think of $\smag$ as providing the multipole-index $j\geq 2$ components of the spin angular momentum \cite{ADH2007}.

\subsection{The far zone; failure of uniformity}

One may define the far zone (if it exists) of a general-relativistic system as the regime in which its geometry is well-approximated by the leading terms in the Bondi asymptotic expansions.  For a realistic radiating system, this will be a zone far enough away from the sources that the gravitational disturbances have resolved into outgoing, transverse waves but not so far away that they begin to encounter other systems or sources of curvature which would distort that behavior.  If the system is not radiating, the lead curvature term will be due to the component $\Psi _2$.

It is important to understand that the Newman--Penrose asymptotic 
expansions describing this regime are {\em not} generally valid uniformly in $u$.\footnote{A simple example is the Schwarzschild solution in a boosted frame.}  In particular, it often happens that the points in the far zone have $r$ increasing as $u$ increases.  
Knowing where the asymptotic expansions are valid is a key 
issue in connecting them to physical interpretations, and it depends on the details of the system at hand.

For example, in the discussion in the previous subsection, I implicitly assumed that the asymptotic form (\ref{agde}) of the geodesic deviation equation held, for each $r$-value under consideration, for a long enough interval of $u$ to stretch from one non-radiating interval $I_1$ to another $I_2$. Since the total interval of retarded time considered is compact, there {\em will} be distant enough $r$-values for this to hold, but just how far out they are will depend on the specifics of the situation.

While this point is always of some concern, it has been possible to ignore for electric memory effects, because, as noted above, there it is at least mathematically self-consistent to assume the the regimes $I_1$ and $I_2$ are stationary.  But in the magnetic cases it cannot be avoided.
This issue will make statements about magnetic effects finicky.

A full resolution of this will depend on thinking more carefully about the physical meaning of the radiation zone.  I wrote above that it was a regime in which the geometry is well-approximated by the leading Bondi--Sachs asymptotics, but to apply this in any problem we must know just what aspect of the geometry is being probed and how good the approximation is required to be.  This will be taken up in Section~VII.

\subsection{Magnetic shear and sources}

We expect a space--time to be determined by suitable initial data for the matter within it and the gravitational degrees of freedom.  What sorts of matter (and what gravitational configurations) would give rise to magnetic shear?

In linearized gravity, the contributions from matter and from gravitational perturbations are independent.  For the gravitational degrees of freedom, Bondi shear at past null infinity $\scrip$ is mapped to  
future null infinity $\scrif$ in a straightforward way.  It is always possible that such data are present.

We can get an idea of how matter generates shear in linearized gravity by looking at th quadrupole terms.\footnote{In fact, the results for higher multipoles can be deduced from these, by taking derivatives and boosts.}
Then the electric and magnetic quadrupole parts of the shear are proportional to the second time derivatives ${\ddot Q}^{\rm el}_{ab}$ and ${\ddot Q}^{\rm mag}_{ab}$ of
corresponding source quadrupoles.  The electric quadrupole $Q^{\rm el}_{ab}$ is familiar as the reduced second mass (or, more properly, energy) moment.  As will be discussed below, the magnetic quadrupole $Q^{\rm mag}_{ab}$ turns out to be proportional to the {\em first} moment of the angular momentum density.  

A quadrupole must vary nonlinearly in time in order to generate shear.  We may expect this in both electric and magnetic cases.  However, for a $u$-{\em independent} shear, the cases are very different.  Such behavior is easy to arrange in the electric case.  (A system which splits into several subsystems with non-zero mutual velocities will have this character --- in fact, this mechanism is fundamental for electric memory effects.)  But for magnetic effects, while it is not hard to imagine mechanisms which for {\em finite} intervals of retarded time will generate CPMS behavior, there is no known way of achieving this for unbounded intervals.  

Finally, I should mention an attempt to suggest an astrophysical situation in which CPMS effects might arise.
If a gravitational wave passes through a volume where there is a chiral fermion density, the electric and magnetic parts of the wave will to some degree interconvert \cite{ADH2016FGG}.\footnote{This is derived by treating the wave as a first-order perturbation on the background space--time.  However, this goes beyond the linearized gravity approximation, as in this case the stress--energy is allowed to respond to the perturbation --- indeed, that is what drives the process.} 
If this also holds in the zero-frequency limit, then a wave-train, which would (if no fermions were present) give rise to a supertranslation, would (after passing though the chiral fermion region) also create a change in $\smag$.  
I used this to sketch a (somewhat elaborate) mechanism by which in principle black holes emitting jets via the Blandford--Znajek process might also acquire $\smag\not=0$ \cite{ADH2018}.

\section{Quadrupole source terms}

In linearized gravity, when we work out the field due to a source,
we integrate the stress-energy $T_{ab}$ against a Green's function.  In the simplest case, if the source were supposed to be a small, featureless, mass, we would idealize $T_{ab}$ by a spatial delta-function.  Of course, point masses are not really admissible in general relativity, and the delta function is not to be taken in any literal sense.  Rather it approximates the effects of a monopole source term as soon as we are a few gravitational radii away.  If the mass is not featureless, it would have multipole moments.  For computations outside the mass, one could idealize their contributions as spatial derivatives of the spatial delta function.
These idealizations are formally simple and computationally powerful; for this we have traded specific knowledge of the physics within the mass.  (In particular, questions about what sources can produce these quadrupoles are not addressable within this formalism.)

One can similarly idealize the effects of distributions of matter in linearized general relativity by point multipoles, and (for multipole index $j\geq 2$) these may be either of electric or magnetic type
(as classified by the shears they produce).
In ref. \cite{ADH2013}, it was shown that stress--energy for a point purely magnetic quadrupole $Q_{ab}(t)$ at the spatial origin is
\begin{equation}\label{se}
  T_{ab} = t^p\epsilon _{pqr(a} \left( -{\dot Q}_{b)}{}^q +t_{b)} Q_s{}^q\nabla ^s
    \right) \nabla ^r\delta ^{(3)}(x)\, .
\end{equation}
Here $Q_{ab}$ is symmetric, trace--free and orthogonal to $t^a$; its time-dependence may be arbitrary; the stress--energy is automatically conserved.\footnote{This stress--energy will not by itself satisfy any energy conditions.  While this is partly due to the singular, distributional, character of the idealization, the main point is that it is not supposed to represent all the matter.  It is simply one multipole component.  It is the full matter distribution which one would take to be subject to energy conditions.}  It is again not to be taken literally; really one should think of a smooth distribution of matter reproducing these quadrupole moments, in the vicinity of the world-line.
It was also shown in ref. \cite{ADH2013} that the magnetic quadrupole from such a distribution can be computed as
\begin{equation}
  Q_{ab} =(4/3)\int {\mathcal L}_{(a}(x_{b)} -tt_{b)})\, d^3 x\, ,
\end{equation}
where ${\mathcal L}_a=\epsilon _{apqr} t^px^q t^cT_c{}^r$ is the angular momentum density.  Thus $Q_{ab}$ can be thought of as a first moment of the angular moment density.  For instance, two parallel hoops of mass $M$ and radius $R$, each orthogonal to the $z$-axis, at
$z=\pm L/2$ and spinning with angular velocities $\pm \omega$,
will give rise to a quadrupole
\begin{eqnarray}
 Q_{ab} &= &{\mathcal Q}\left[
   \begin{array}{cccc}
   0&&&\\ &-1&&\\ &&-1&\\ &&&2
   \end{array}\right]\, ,\label{hoopsquad}\\
   {\mathcal Q}&=&(2/3) M\omega R^2 L\label{qscalar}
\end{eqnarray}
in standard Cartesian coordinates (blank places are zero).

It is straightforward to compute from eq. (\ref{se}) the retarded linearized metric perturbation $h_{ab}$ in the de Donder gauge.  One finds
\begin{eqnarray}\label{metpert}
h_{ab}&=&4Gt^p\epsilon _{pqr(a}x^q\left\{
       r^{-2}{\ddot Q}_{b)}{}^r +r^{-3}{\dot Q}_{b)}{}^r\right.\nonumber\\
       &&\left. -t_{b)}x^j\left[ r^{-3}{\ddot Q}_j{}^r +
    3 r^{-4}{\dot Q}_j{}^r +3r^{-5}Q_j{}^r\right]\right\}\, . \qquad
\end{eqnarray}
Here $Q_{ab}$ and its derivatives are evaluated at the retarded time $t-r$.

To give an invariant account of scattering for massless particles, we must pass to the Bondi--Sachs gauge.  To do this, first note that the retarded time $u=t-r$ remains a null coordinate for the perturbed metric $g_{ab}=\eta _{ab} +h_{ab}$.  The affinely-parameterized null geodesic congruence ruling the $u=\const$ hypersurfaces will be $l^a-h^a{}_b l^b$, and thus the perturbation 
$\delta\gamma_{\rm B} ^a =-\int h^a{}_b l^b\, dr$ (the integral being taken along the geodesics)
of these geodesics relative to those for the Minkowski background can be computed:
\begin{eqnarray}
     \delta \gamma_{\rm B} ^a=   -2Gt^p\epsilon_{pqr}{}^a l^q l^b
        \left[ 2r^{-1}{\dot Q}_b{}^r +(3/2) r^{-2} Q_b{}^r \right]
        \, .\qquad
\end{eqnarray}    
(The subscript B is for {\em Bondi congruence}.)
This shows that the perturbed outgoing geodesics approach the unperturbed ones as $r\to\infty$.  One then sees by inspection that the fall-off of the perturbation along the $r=\const$ cross-sections of the $u=\const$ hypersurfaces has the requisite asymptotics.
Because the Bondi--Sachs coordinatization at $\scrif$ itself agrees with the Minkowskian one, for the questions investigated below it will not be necessary to take the gauge change to this system into account  
(although for more delicate questions this would be relevant), and the form (\ref{metpert}) will be used.
However, the formulas for the gauge change will be given for completeness. 

Suppose we wish to label a point $x^a$ by its Bondi--Sachs coordinates.  We have seen that the retarded time coordinate $u$ is unaffected by the metric perturbation.  The new affine coordinate will be $r+\delta r$, where $(l^a-h^a{}_bl^b)\nabla _a(r+\delta r) =0$ or $\delta r = \int _\infty^r (h^{ab}l_a\nabla _br)\, dr$ along the outward null geodesics.  Finally, we wish to label each point by the angles corresponding to the point on $S^2$ determined by the tangent to the null geodesic outwards from the point, evaluated asymptotically as we approach $\scrif$.  The null geodesic outwards from $x^a$ will be
\begin{eqnarray} 
  \gamma ^a(s) =x^a +s(l^a-h^a{}_b(x)\, l^b) +\delta\gamma_{\rm B}^a\, .
\end{eqnarray}
Because of the fall-off of $\delta\gamma_{\rm B}^a$,
the angles for the asymptotic tangent are those for the point $l^a-h^a{}_b(x) \, l^b$ on the sphere.

The explicit form of the metric allows one to work out the full scattering theory in terms of integrals of $Q_{ab}$ and its derivatives (times certain functions) along the Minkowskian geodesics.  However, the formulas are lengthy, and in this paper I shall just focus on examples of special interest.

We may read off from eq. (\ref{metpert}) the asymptotic form of the curvature near $\scrif$; it is
\begin{eqnarray}
  R_{abcd}&=&\nabla _c\nabla _{[a}h_{b]d}
     -\nabla _d\nabla _{[a}h_{b]c}\label{curv}\nonumber\\  
  &\simeq& 4r^{-1} Gt^p\epsilon _{pqr[d}l^ql_{c]}
    Q^{(4)}_{[b}{}^r l_{a]} \nonumber\\
    &&+4r^{-1} Gt^p\epsilon _{pqr[b}l^ql_{a]}
    Q^{(4)}_{[d}{}^r l_{c]}
      \, ,
\end{eqnarray}
where the superscript $(4)$ indicates the fourth derivative.
From this we have
\begin{eqnarray}
\Psi _4^0 &=& (i/2)G Q^{(4)}_{ab}{\overline m}^a{\overline m}^b\, .
\end{eqnarray}
Eq. (\ref{psifour}) then gives the Bondi shear:
\begin{eqnarray}
  \sigma = (i/2)G{\ddot Q}_{ab}m^a m^b\, .
\end{eqnarray}    
We see that in order to get a non-trivial but $u$-independent magnetic shear, we must have $Q_{ab}$ depend quadratically on $u$.  While this can reasonably be maintained for finite intervals, 
it is not at all clear if a mechanical configuration can be devised doing so indefinitely.\footnote{An indefinite linear growth is easy to achieve, for example by using the hoop model but allowing the hoops to move away from each other on the $z$-axis.}
One might be tempted to argue that an unbounded quadratic growth of $Q_{ab}$ is evidently unphysical, based on the consequent growth of the stress--energy (\ref{se}) and the metric (\ref{metpert}).  While there is a sense in which this is true, it is not a strong sense.  What we really learn by inspecting these formulas is that the following three assumptions are not simultaneously compatible:  that we may approximate the source as a point quadrupole; that we may apply linearized gravity; and that the quadrupole grows indefinitely in time.
This certainly does place restrictions on the regimes in which the formulas are applicable, but it does not invalidate them wholesale.

It may help to think about the case of electric quadrupoles.
Recall that a system breaking into relatively moving subsystems will give rise to an electric quadrupole growing quadratically with time.  At any finite time, on a large enough spatial scale --- much larger than the separations between the subsystems --- one can approximate the system by point multipoles.  So on large enough scales electric analogs of eqs. (\ref{se}), (\ref{metpert}) will apply.  Those scales will moreover grow with the passage of time, since the separations between the subsystems is growing.  

In other words, by idealizing the problem by using a point quadrupole as a source, we are able to get what appear to be explicit solutions.  But those solutions are {\em only} valid on large enough scales that the idealization is a good one.  Just what those scales are depends on the particular system, and cannot be read off simply from the idealization.  If the magnetic quadrupole does grow with time, then formula (\ref{metpert}) certainly breaks down at any given $r$ after a long enough time.  But whether this signals that it is impossible to maintain a value of $\smag$ indefinitely, or rather that for a realistic matter distribution the metric in finite regimes is more complex, is impossible to say without detailed analysis of realistic matter.

\section{Initially comoving geodesics}

In this section I will consider the effects of the quadrupole on geodesics which are initially comoving with the source.  (The quadrupole will be assumed to vanish sufficiently far in the past.)  While this is a rather special situation, it is the zeroth-order approximation to the more general case of non-relativistic motion, and there are interesting things to learn from it.

The first subsection derives the formulas for the geodesics.

The second subsection will compare this case with the general arguments about radiation memory from Section~II.  Recall that those arguments, derived in the far zone, gave the change in displacement (\ref{memeq}) of nearby particles in terms of the memory tensor.  We will see how this comes up in our case, but we will also see that the identification of the far zone is somewhat involved, limiting where the simple form (\ref{memeq}) applies.  

The third subsection establishes the Lenz's-law-type result, that in the induction zone a stationary quadrupole acts on nearby test particles in a way as to induce an opposing quadrupole.

The fourth subsection is concerned with induction-zone memory.  It is there that a possible laboratory effect is identified.  Various potential astrophysical effects are also considered, but no promising candidates are found.

\subsection{The geodesics}

The geodesics in the past have the form
\begin{eqnarray}
\gamma_{\rm p}^a(s) =b^a +st^a\, ,
\end{eqnarray}
where we may take the impact vector $b^a$ purely spatial.  Then because $h_{ac}{\dot\gamma}^a{\dot\gamma}^c=0$, there is a first integral of the geodesic equation and we find
the perturbations of the velocities are
\begin{eqnarray}\label{velpert}
\delta{\dot\gamma}^a(s) &=& 2G t^p\epsilon _{pqr}{}^ab^qb^j
  [ b^{-3}{\ddot Q}_j{}^r + 3b^{-4}{\dot Q}_j{}^r + 3b^{-5}{Q}_j{}^r
  ]\, ,\qquad
\end{eqnarray}
where $b=\sqrt{-b_ab^a}$ and the quadrupole and its derivatives are evaluated at the retarded time $s-b$.  
We see that in any fixed spatial region it is inconsistent to assume simultaneously linearized gravity, a point quadrupole, and indefinite quadratic growth of $Q_{ab}(t)$.

It is worth noting that the form (\ref{velpert}) shows the velocity is orthogonal to $b^a$, so (to the extent this linearized treatment is valid) the perturbed geodesics will remain on the coordinate spheres $r=b$.  On the other hand, simply computing the trajectories in the linearized approximation we find
\begin{eqnarray}\label{dgameq}
\delta\gamma ^a &=&  2G t^p\epsilon _{pqr}{}^ab^qb^j
  \Bigl[ b^{-3}{\dot Q}_j{}^r + 3b^{-4}{Q}_j{}^r \nonumber\\
  &&
  + 3b^{-5}\int _{-\infty}^{s-|b|}{Q}_j{}^r(\acute s )
 \, d\acute s \Bigr]\, .
\end{eqnarray}
The last two terms indicate that the linearized approximation cannot be valid uniformly in time if $Q_{ab}$ is allowed to increase indefinitely, or indeed even reach a steady, non-zero, state.  This is not surprising, as even in Newtonian mechanics, the effects of a small force acting for a long enough time usually accumulate and pass beyond perturbation theory.

\subsection{Consequences for magnetic radiation memory}

Radiation memory is usually considered to be the change in relative displacement of nearby test masses caused by passage of gravitational radiation.  We saw in Section~II that in the radiation zone, taking advantage of the Bondi--Sachs--Newman--Penrose formalism led to a simple formula (\ref{memeq}) for this, with the memory tensor expressed in terms of the change in shear.  
It is instructive to examine radiation memory in the present case.

For magnetic quadrupole sources in linearized gravity, we may read off the geodesic deviation by differentiating eq. (\ref{dgameq}) with respect to $b^a$ in the direction $w^a$ of the separation of two geodesics.  That is, the change in this separation due to the quadrupole is
\begin{eqnarray}\label{dweq}
  \delta w^a &=& w^p\frac{\partial}{\partial b^p}\delta\gamma ^a\, .
\end{eqnarray}
Formally, the leading (long-distance) behavior of this is the term
\begin{eqnarray}\label{ftdw}
  2Gt^p\epsilon _{pqr}{}^a b^qb^j b^{-4} w^sb_s{\ddot Q}_j{}^r\, ,
\end{eqnarray}
corresponding to the formula (\ref{memeq}) for the memory in the radiation zone.  

However, one must be careful about the sense in which this really is the dominant term.  It will clearly be so at fixed $u$ for sufficiently large $b$.  On the other hand, suppose in some regime
$\smag$ is constant, so $Q_j{}^r$ grows quadratically with $u$.  Then for fixed $b^a$, the terms in (\ref{dweq}) other than (\ref{ftdw}) will grow with $u$ and entually
overwhelm (\ref{ftdw}).  (And at some point the linearized approximation itself will break down.)
This is an example of the the non-uniform dependence of the far zone on $u$.  

This does not mean magnetic radiation memory, as defined by eq. (\ref{memeq}), is impossible.  It does however mean that there are serious restrictions on when it can apply.  One needs to know that the test particles are indeed in the radiation zone, and this will be more problematic than the electric case.  I will return to this in Section~VII.

\subsection{Stationary states and Lenz's law}

In the case where the quadrupole settles down and becomes time-independent, the equation for its contribution to the velocity perturbation (\ref{velpert}) reduces to
\begin{eqnarray}\label{newvel}
\delta{\dot\gamma}^a(s) &=& 6G b^{-5}t^p\epsilon _{pqr}{}^ab^qb^j
 {Q}_j{}^r  \, .
\end{eqnarray}
In almost all circumstances, there will be other contributions as well (for instance, monopole terms).  But the mathematical structure of eq. (\ref{newvel}) is so remarkable that a brief comment is in order.

Equation (\ref{newvel}) is a very interesting system of equations for the spatial coordinate vector ${\bf x}$.  As noted above, it preserves the coordinate radius.  There is also another constant of motion, which is $Q_{ab}x^ax^b$.  This means that the particles' trajectories are the intersections of these quadratic surfaces.  Generically, these will be quartics, and the trajectories can be computed in terms of elliptic integrals by choosing the coordinate axes to be the principle axes for $Q_{ab}$. 

I now return to thinking of eq. (\ref{newvel}) simply as giving the quadrupole's contribution to the equation of motion.  
For simplicity, let us look at the case where the quadrupole has two equal eigenvalues.  Then it will have the same form (\ref{hoopsquad}) as that for the two counter-rotating hoops, and this expression will be retained.
In this case, we find eq. (\ref{newvel}) becomes, in three-vector form,
\begin{eqnarray}
\delta\dot{\bf x} &=& 
        -18 G{\mathcal Q}b^{-5} ({\bf k}\cdot{\bf x})
  ({\bf k}\times{\bf x})\, ,
\end{eqnarray}
where ${\bf k}$ is the unit vector in the $+z$ direction.
Note that the induced motion in this case will also consist of revolutions about the $z$-axis, but their sense will {\em oppose} those of the original hoops.  Therefore, at least in this case, the induced motion will tend to generate, from a distribution of free particles, a quadrupole {\em opposed} to the initial one.  This has the flavor of Lenz's law, although here the particles' motion is induced by $Q_{ab}$ (and not some time-derivative of that).  Note that this argument does not depend on integrating the equation of motion (and hence on questions of how long it will be before nonlinearities accumulate).

\subsection{Induction-zone memory}

Now let us consider what happens to initially comoving geodesics when a non-zero quadrupole is present for a finite amount of time only.  In this case, there will be a net displacement
\begin{eqnarray}\label{disp}
\delta \gamma ^a &=&6Gt^p b^{-5} \epsilon _{pqr}{}^a b^q b^j
  Q^{(-1)}{}_j{}^r\, ,
\end{eqnarray}
where
\begin{eqnarray}
Q^{(-1)}{}_j{}^r=\int_{-\infty}^\infty Q_j{}^r(s)\, ds\, .
\end{eqnarray}

I will suppose for simplicity that we are in the axial case
\begin{eqnarray}
Q^{(-1)}{}_{ab} ={\mathcal Q}^{(-1)}\left[\begin{array}{cccc}
0&&&\\
   &-1&&\\ &&-1&\\&&&2\end{array}\right]
\end{eqnarray}
with respect to the coordinate axes, where ${\mathcal Q}^{(-1)}$ is a scalar 
(blank places are zeroes).  Then the displacement (\ref{disp}) becomes in ordinary vector notation
\begin{eqnarray}\label{dexeq}
  \delta{\bf x} &=&-18 G r^{-5}{\mathcal Q}^{(-1)} ({\bf k}\cdot {\bf x})
   ( {\bf k}\times {\bf x})\, .
\end{eqnarray}   

Some care about the physical interpretation of this formula is in order.  Here ${\bf x}$ form three of the coordinates to the geodesic, and in general relativity (even in the linearized theory) coordinates do not {\em a priori} have physical meaning.  However, we have been supposing that the quadrupole vanishes except for a finite range of times, and that means that the metric will be Minkowskian except where the source influences it in accordance to Huygens's principle.  So both before and after the influence of the source, we have a clear physical interpretation of ${\bf x}$ not simply as coordinates but as the spatial part of the Minkowksian position vector of the source relative to the central world-line.  

At this point, we have argued that we have Minkowskian regimes before and after the quadrupole is present, and also that we have spatial coordinate vectors ${\bf x}$ in each of those.  We must however specify how to compare the two regimes, and, because there is curvature in the interim, this is a non-trivial point.  One might first think of simply parallel-transporting along the central world-line, but because we have used an idealized point quadrupole the metric becomes singular there.

The most invariant thing to do is to use the asymptotic structure to compare the regimes, that is, to identify them by identifying their asymptotically constant vector fields near $\scrif$.
It follows from the analysis of ref.
 \cite{ADH2014} that, because the radiation is confined to a compact interval of retarded time, this simply amounts to identifying the components with respect to Cartesian coordinates in the two regimes.
So the coordinate difference $\delta{\bf x}$ of eq. (\ref{dexeq}) has an invariant interpretation.

The effect (\ref{dexeq}) has what is sometimes called unnatural parity, in the following sense:  Under the antipodal map ${\bf x}\to -{\bf x}$, the quantity $\delta {\bf x}$ is unchanged.  Yet this is a displacement of ${\bf x}$, and, the image of a displacement under the antipodal map is the {\em opposite} displacement.  Thus the operations of forming the displacement and applying the antipodal map {\em anti}commute.

\subsubsection{A potential laboratory effect}

If we know the central world-line of the source, and we have information about the initial segment of the geodesic, then $\delta{\bf x}$ is interpretable as the change in coordinate relative to the source.  It is conceivable that effects like this could be measured in laboratories.  Suppose, for example, the two-hoop source were placed near one test-mass of a laser interferometer.  In general, the interferometer will measure the change in position between the two masses, which is rather more complicated than $\delta{\bf x}$.  However, as noted above, before and after the non-zero values of the quadrupole, the positional measurements are those of Minkowski space.  In these cases, because the second mass is so far away that the effect of the quadrupole on it is negligible, we may interpret $\delta{\bf x}$ as the change in position.

Choosing $M=10^6\, {\rm g}$, $R= 3\times 10^2\, {\rm cm}$, $\omega = 10^3\, {\rm s}^{-1}$, $L=10^2\, {\rm cm}$, a distance $r=10^3\, {\rm cm}$ and a run-time $T$, we should have
\begin{eqnarray}
  \| \delta{\bf x} \| &\sim&  \left( 3\times 10^{-13} \, {\rm cm}\right)
  \left(\frac{T}{1\, {\rm y}}\right) \sin\theta\cos\theta 
   \, .
\end{eqnarray}
While the prefactor looks encouraging, current gravitational-wave interferometers are designed to measure oscillatory effects, and what we are considering here would show up as a zero-frequency, linear, drift.  The detectability of this would depend on the the temporal stability of the interferometer.  In this connection, see the suggestion of Lasky et al. \cite{LTLB2016} for a statistical approach to accumulating interferometric measurements for memory.

\subsubsection{Potential astrophysical effects}

Now let us consider possible astrophysical effects.  We suppose there is an object which might have been active at times as a magnetic quadrupole source, and that we can observe matter in its vicinity; we seek possible memory effects.

The chief issue in this case is to find, in the vicinity of the source, distributions of matter whose original states might be known or plausibly inferred to the necessary accuracy.  
If, for instance, we had reason to think that some mechanism had formed filaments longitudinally with respect to the axis of the quadrupole, then eq. (\ref{dexeq}) would imply the filaments would acquire S-shapes as a result of the quadrupole's action, bulging azimuthally one way in one hemisphere and the other way in the other.

The effect just described relied on a plausible hypothesis about structure (filaments) extending over substantial angles on the sphere around the quadrupole source.  We may also consider hypotheses about structure on smaller scales.  If we knew the
separation $\Delta {\bf x}$ between two nearby geodesics carrying particles, it would change, after the effects of the quadrupole, by an amount
\begin{eqnarray}\label{pdist}
\delta\Delta {\bf x}&=& -18G{\mathcal Q}^{(-1)} r^{-5}\left[
   ({\bf k}\cdot \Delta{\bf x})({\bf k}\times{\bf x})\right.\nonumber\\
   &&\left.
  +({\bf k}\cdot{\bf x})({\bf k}\times\Delta{\bf x})\right.\nonumber\\
  &&\left.
  -5r^{-2}({\bf x}\cdot{\Delta{\bf x}})({\bf k}\cdot{\bf x})({\bf k}\times {\bf x})\right]\, .
\end{eqnarray}  
This is a first-order differential effect.

The general second-order differential effect is to shear a geodesic congruence (physically, a distribution of particles initially comoving with the source).
To see this, let us think of an initially spherical distribution of particles in the neighborhood of a geodesic.  We may then consider $\Delta{\bf x}$ to be a random variable where the sample space is the set of these particles.
We will assume
\begin{eqnarray}
  \langle \Delta x^j\rangle &=&0\\
  \langle \Delta x^j \Delta x^k \rangle &=& (1/3)(\Delta x)_{\rm rms}^2
  \varepsilon^{jk}\, ,
\end{eqnarray}
where the brackets $\langle\cdots\rangle$ denote statistical average
and $\varepsilon^{jk}$ is the Euclidean metric.  
Then 
\begin{eqnarray}
  \langle \Delta x^j +\delta\Delta x^j \rangle &=&0
\end{eqnarray}
and (to first order)
\begin{eqnarray}\label{fluceq}
  &&\langle (\Delta x^j +\delta\Delta x^j) (\Delta x^k +\delta\Delta x^k) 
\rangle  \nonumber\\
& &\qquad = (1/3)(\Delta x)_{\rm rms}^2
  \varepsilon^{jk} -6 (\Delta x)_{\rm rms}^2G{\mathcal Q}^{(-1)} r^{-7}\times\nonumber\\
  &&
  \qquad \left[
  \begin{array}{ccc}
    10xyz  & 5(y^2-x^2) z   &  5yz^2 -r^2y \\
    5(y^2-x^2)z   & -10xyz   & -5xz^2 +r^2x\\
    5yz^2 -r^2y & -5xz^2 +r^2x & 0
    \end{array}\right]\, .\qquad
\end{eqnarray}   
(In case the the asymmetry in $x$ and $y$ appears odd, it should be remembered that this matrix really refers to components of tangent vectors at a point $(x,y,z)$.)  
The most important point is that this is trace-free, and therefore the local density --- which might have been a relatively straightforward thing to try to measure --- is unchanged by quadrupole's action.  The deformation is pure shear.

To understand the formula (\ref{fluceq}),
we may by rotational symmetry restrict attention to the meridian $y=0$, $x\geq 0$.
Then the correction term is
\begin{eqnarray}\label{mersh}
&&-6 (\Delta x)_{\rm rms}^2G{\mathcal Q}^{(-1)} r^{-4}\nonumber\\
&&\qquad \times
  \left[
            \begin{array}{ccc} 
          0&-5x^2z & 0\\
          -5x^2 z& 0&x(r^2-5z^2)\\
          0& x(r^2-5z^2)&0
          \end{array}
          \right]\, .\qquad
\end{eqnarray}  
The zeroes on the diagonal mean that, on our meridian, the quadrupole contributes no change to the distribution's dimensions along the coordinate axes.  Expressing this invariantly, we may say that
at any point, the 
dimensions in the $z$-direction, and also the azimuthal direction and the direction of constant latitude, receive no changes from the quadrupole's action.  On the other hand, the eigenvectors of the matrix above will determine the principal axes for the shear.  

It is not hard to see from eq. (\ref{mersh}) that one principal axis of the shear is 
\begin{eqnarray}
  \propto\left[\begin{array}{c} x(r^2-5z^2)\\ 0\\ 5x^2z\end{array}\right]\, ,
\end{eqnarray}  
in the longitudinal plane, with corresponding eigenvalue zero, 
so no shearing occurs in this direction.  While the detailed forms of the other principal axes are complicated, it turns out that if the shear tensor is restricted to the tangent plane of the sphere --- that is, if we ask for the effects of the shear projected orthogonal to the radial direction --- the results are simple.
One can check
(again from eq. (\ref{mersh})) that the principal axes of the projected tensor are at $\pm \pi/4$ relative to the latitude--longitude lines, and
the angular dependence of the eigenvalues is $\sim \sin\theta$.
So if we were lucky enough to have objects we could plausibly assume had initially been round distributed about the candidate quadrupole source, and if we were able to measure their strains, we would have a straightforward check of whether these could have been produced by the quadrupole.

\section{A red-shift effect}

Because a persistent magnetic shear will give rise to a curvature term with the same radial fall-off as the Newtonian one, one would like to look for a scattering effect due to magnetic shear which is comparable to the Newtonian one, that is, falls off as the reciprocal of the impact parameter (in the large-impact-parameter limit).  
It is not obvious just what effect might have this character, for the effects depend  very much on the detailed form of the curvature, and the quadrupole dependence makes it hard to see just which effects might accumulate.  Moreover, as  noted previously, the space--time cannot be stationary, making it still harder to guess what the effects might be without detailed computations.

It turns out that the red-shift does fall off as the reciprocal of the impact parameter (in the limit that this parameter is large).
Consider a null geodesic in Minkowski space
\begin{eqnarray}
  \gamma ^a_{\rm p}(s) = b^a +u_0t^a +sL^a\, ,
\end{eqnarray}
where $L\cdot t = 1$, $L\cdot b =0$, $t\cdot b =0$.
From the geodesic equation, the perturbation $\delta\gamma^a(s)$ of this due to the metric perturbation $h_{ab}$ satisfies
\begin{eqnarray}
\delta{\ddot\gamma}^a &=& -{\dot\gamma}^b\nabla _b h_c{}^a{\dot\gamma}^c +(1/2){\dot\gamma}^b{\dot\gamma}^c\nabla ^ah_{bc}\, .
\end{eqnarray}
Integrating this in order to find the scattering, the first term on the right drops out, and, using the fact that at zeroth order ${\dot\gamma}^a =L^a$ is constant in Minkowski space, we have
\begin{eqnarray}
{\dot\gamma}^a\Bigr|_{s=-\infty}^{+\infty}
  &=&(1/2)\int _{-\infty}^\infty\nabla ^ah_{bc} L^b L^c\, ds\, .
\end{eqnarray}

The net change in the temporal component of this will be $-z$, the negative of the red-shift.\footnote{So here $z$ does not stand for a coordinate.}  Explicitly
\begin{eqnarray}
  z &=&-(1/2)\int 4Gt^pL^a  \epsilon _{pqra}b^q
    \left\{    
    L^b [r^{-2} Q^{(3)}_b{}^r + r^{-3}{\ddot Q}_b{}^r]
      \right.\nonumber\\
      &&\left. -(b^j+sL^j) [r^{-3}Q^{(3)\, r}_j +3r^{-4}{\ddot Q}_j^r
      +3r^{-5}{\dot Q}_j^r]\right\}\, ds\, .\qquad
\end{eqnarray}     
It is straightforward to estimate this for large $b$ (see the appendix).  One finds
\begin{eqnarray}\label{rs}
z&\simeq &-4Gt^pL^ab^q\epsilon _{pqra}b^{-2}L^b{\ddot Q}_b{}^r(u_0)
    \, .
\end{eqnarray}  

We see that the red-shift (\ref{rs}) does indeed fall off as the reciprocal $b^{-1}$ of the impact parameter.  

Perhaps the next most striking feature of the formula is that it depends on the quadrupole's value only at $u_0$.  This is a retardation effect.  
The computation is given in the appendix, but the reason for the result is this.
The geodesic equation depends on the quadrupole and its derivatives evaluated at the retarded time $u$ of the point in question.  We have
$u= u_0 + s -\sqrt{b^2+s^2}$.  We are looking at the limit of large $b$ (large, in particular, compared to the duration of the action of the quadrupole).  We will need to have $s$ at least of the order of $b$ in order to access the retarded times for which the quadrupole may be non-zero.  But in this regime $u$ approaches $u_0$ very rapidly (in terms of the scale $b$).
In contrast to memory effects, one might say that the red-shift has an {\em amnesiac} character --- it depends (in the limit $b\to\infty$) only on the last value of ${\ddot Q}_{ab}$ accessible to the scattered ray. 

In particular, we see that the red-shift requires ${\ddot Q}_{ab}(u_0)\not=0$.  This means that it does depend on generating a magnetic shear; on the other hand, it does {\em not} depend on that shear persisting indefinitely.  

In three-vector notation, the red-shift is
\begin{eqnarray}\label{zapp}
  z &=&4G b^{-2} (({\bf b}\times {\bf v})\cdot {\bf L})\, ,
\end{eqnarray}
where ${\bf b}$ and ${\bf L}$ are the spatial parts of $b^a$  and $L^a$, and ${\bf v}$ is the spatial part of ${\ddot Q}_b{}^r L^b$.
The red-shift reverses sign under the inversion of the spatial parts ${\bf L}\to -{\bf L}$, ${\bf b}\to -{\bf b}$, so it has odd parity in a straightforward sense.

For celestial sources, in the simplest cases ${\bf L}$ is fixed as the unit vector in the direction from the source to us, but ${\bf b}$ will be the impact vector, in the plane of the sky.  The red-shift will vary as the cosine of the angle ${\bf b}$ makes with ${\bf v}\times {\bf L}$.  
It should be noted, however, that such a sinusoidal red-shift would also be produced from a boosted Schwarzschild solution.\footnote{In linearized gravity, the deflection of light can be written invariantly as $-4GMb^a/b^2$, where $b^a$ is the impact vector with respect to the frame defined by the source --- it is orthogonal to the vector $t^a$ defining the source's frame, as well as the null tangent $L^a$.  If the null geodesic is held fixed to zeroth order but the source is boosted so the frame vector becomes ${\acute t}^a$, then the impact vector with respect to this frame will be $b^a -L^a({\acute t}\cdot b)/({\acute t}\cdot L)$ (and the impact parameter will be unchanged).  The last term will have a timlike component, and accordingly there will be a red-shift $(4GM/b){\acute t}\cdot b/{\acute t}\cdot L$.}
Thus one would also need a measurement of the source's velocity in order to distinguish the effect of the magnetic quadrupole from that of a boosted monopole.

The boosted monopole creates a dipole of electric type, which gives the sinusoidal contribution to the redshift.  One would like to know also what the effects of higher electric multipoles are, and what is required observationally to distinguish them from the magnetic effects of interest here.  Even for quadrupoles, this is a substantial problem, and will be investigated elsewhere.

If we were lucky enough to have a lensing mass between us and the source, we might be able to make measurements for several different values of ${\bf L}$, and these would help a great deal in identifying the gravitational field.  For instance, 
the magnetic quadrupole effect depends quadratically on ${\bf L}$, whereas the boosted monopole effect is linear (to lowest order in velocity).\footnote{As follows from the formula at the end of the previous footnote.}

This red-shift is presumably a better candidate for an observable effect than were the memory ones, partly in that it is a longer-range effect and partly that the measurements involved would be much less fussy.  On the other hand, it does require catching the quadrupole source while it is varying quadratically.

\section{Far zone for a CPMS quadrupole}

I defined the far zone as a regime in which the geometry is well-approximated by the leading terms in the Bondi--Sachs expansion.  Although this is a good intuitive beginning, it does need some refinement.  The issue is that many quantities of physical interest are nonlocal; for these we typically need to know that some integrals of geometric quantities over extended sets are suitably controlled.  The sets on which this will hold will depend on the physical quantities of interest, so
really one should speak of what regime should be considered the far zone for a given sort of measurement.

The most important class of observables will be ones derived from the trajectories of test particles, and if we are interested in results which stabilize as the intervals of measurement increase, then we must take complete geodesics.  This will be done here, in the case of geodesic scattering from a CPMS quadrupole in linearized gravity.

The first task will be to examine the metric perturbation along geodesics in the background space--time (Minkowski space), and make sure this is controlled.  After that, the scattering itself, which requires an integral over the geodesic, will be computed.

The zeroth-order geodesic, in the background Minkowski space--time, will be written
\begin{eqnarray}\label{zerogeod}
  \gamma _0^a (s) &=& b^a +u_0 t^a +s (t^a\cosh\xi + \z ^a\sinh\xi )\, ,
\end{eqnarray}
where $\z ^a$ is a unit spacelike vector, the vectors
$b^a$, $t^a$, $\z ^a$ are mutually orthogonal, and $\xi$ is the rapidity.  As before, I will write $b=\sqrt{-b_ab^a}$ for the impact parameter.  
Along the geodesic, the coordinates are given by
\begin{eqnarray}
  r&=&\sqrt{b^2+s^2\sinh^2\xi}\label{ceqa}\\
  u&=& u_0+s\cosh\xi -\sqrt{b^2+s^2\sinh^2\xi}\label{ceqb}\\
  l^a &=& t^a +(b^a +s(\sinh\xi )\z ^a)/r\, .\label{ceqc}
\end{eqnarray}

Now let us turn to the metric perturbation.  Suppose for simplicity the quadrupole has a purely quadratic dependence on $u$, so $Q_{ab} =(1/2)u^2 {\ddot Q}_{ab}$ for some constant ${\ddot Q}_{ab}$.  Inspection of the formula (\ref{metpert}) for the metric perturbation shows it can be written as a sum of terms of the form $({\ddot Q}_{ab}/r)(u/r)^n$ for $n=0,1,2$ contracted with tensors whose components are of order at most unity.

Along the geodesic, the factor ${\ddot Q}_{ab}/r$ will be bounded by $|\lambda|/b$, where $\lambda$ is the eigenvalue of ${\ddot Q}_{ab}$ of the largest magnitude.  A short computation shows that the factor $u/r$ has limiting values
\begin{equation}
  \pm |\coth\xi| -1
\end{equation}  
as $s\to\pm\infty$, and (if $u_0\not=0$) a single local extremum at $s=(b^2/u_0)\coth\xi \csch\xi$, with value
\begin{equation}
  \sgn (u_0) \sqrt{ (u_0/b)^2 +\coth ^2\xi} -1\, .
\end{equation}  

We may now see how the metric perturbation along the geodesic can be controlled.  We take $b$ large enough so the ratio $\lambda /b$ of the largest-magnitude eigenvalue of ${\ddot Q}_{ab}$ to the impact parameter will be small.  The other factors will be controlled by requiring the rapidity $\xi$ be at least moderate, and the ratio $|u_0/b|$ be at most moderate.  
Then the perturbation will be uniformly small over the geodesic.

It turns out that these restrictions are also enough to control the scattering.
The computation is lengthy but straightforward; I will give only the solution and only the limiting form for $b\gg |u_0|$.  It is
\begin{eqnarray}\label{mscatt}
 \Delta{\dot\gamma}^c &=&-2G (\cosh\xi)(\csch^2\xi)
   t^p\epsilon _{pqra}\z^a{\ddot Q}_j{}^r\nonumber\\
   &&\cdot
     [b^{-2}\Pi ^{qc}b^j +b^{-2}\Pi^{jc}b^q +2b^{-4}b^qb^cb^j]\, ,
\end{eqnarray}
where
\begin{eqnarray}
  \Pi ^q{}_c =\delta ^q{}_c-t^qt_c+\z^q\z_c     
\end{eqnarray}
is projection orthogonal to $t^a$ and $\z^a$.  Recall that $\xi$ is at least moderate here;
the divergence of the scattering (\ref{mscatt}) as $\xi \to 0$ is a failure of the linearized approximation to hold good for the perturbation over the infinite range of $s$ values.\footnote{While the formula (\ref{mscatt}) diverges for large $\xi$, this is because with our normalization the original vector ${\dot\gamma}^a$ does, too (eq. (\ref{zerogeod})).  In fact, the {\em relative} change of ${\dot\gamma}^a$ vanishes in this limit.  It is worth remarking that taking this limit and rescaling does {\em not} reproduce the red-shift formula (\ref{zapp}) for light because it does not also incorporate an appropriate scaling for $u_0$ or $b$.}

Briefly, we may say that we have identified a far zone, suitable for analyzing test-particle scattering, of relativistic (since $\xi$ should be at least moderate) geodesics with $|u_0/b|\ll 1$.

\section{Discussion}

Magnetic gravitational effects are inherently non-Newtonian, general-relativistic features.  
Generically, one expects these degrees of freedom to be non-trivial, and in particular magnetic shear to be present.  Yet we have little understanding of precisely how this should arise, and what its consequences might be.  

The aims of this paper have been to clarify some aspects of the theoretical bases for magnetic shear (particularly as they relate to possible radiation memory effects), and to explore some possible experimental or observational consequences of magnetic effects.  I will review here where some aspects of this stand.

\subsection{Sources}

We do not have a good understanding of the sources of magnetic shear, and this remains a major problem.
This paper has not suggested any new sources of magnetic effects.  On the other hand, it has investigated the fields
due 
to arbitrarily varying quadrupole point sources in linearized gravity.  (Since any magnetic shear given at $\scrif$ can be realized as a superposition of quadrupole contributions and their derivatives, this provides the basis for a general treatment of the linearized theory.)  The computation was done in a de Donder gauge, but the transition to a Bondi gauge was found as well, and the required change was quite mild.

In linearized gravity, we saw that magnetic quadrupoles were associated with first moments of the angular momentum density.  This suggests that we look for nonlinear effects where there are separated contributions to the total angular momentum, for instance, a binary black hole system with opposing spins.  Although it would be hard to detect $\smag$ directly in data from current numerical simulations, one has a hope of recovering it by integrating the magnetic part of $\Psi _4^0$ (see eq. (\ref{psifour})).

It would also be natural to explore the potential for effects due to explicitly chiral matter.  A sketch of how this might come about was suggested in ref.~\cite{ADH2018}.

\subsection{CPMS regimes}

If a gravitational system relaxes to a point where it is not emitting radiation, one would {\em a priori} expect it to retain some magnetic shear, that is, to be in what I have called a CPMS state.  
It is possible that such regimes are indeed the natural end-states for many gravitational systems, but there are several reasons for thinking the situation is more complicated.

One of these comes from work on black holes.  An end-state black hole with magnetic shear would be a strong violation of the No-Hair Conjecture.  
There is a fairly good argument that such a violation is impossible, but it is worth being careful about just what the argument is, because we shall see that some of the relevant physics has not been explored.

As is well known, there are many studies of perturbations of black holes, and a great body of evidence that perturbed holes ``ring down'' via quasi-normal modes to Kerr--Newman states.  However, most of these studies take the initial data for the perturbation to have compact support.  For CPMS regimes, though, this restriction is inappropriate --- the perturbations in question would have zero-frequency components.  The zero-frequency case {\em was} considered by Teukolsky \cite{Teukolsky1972}, who gave arguments that such perturbations were not be stable, but beyond that we have very little knowledge of the dynamics of the situation.  We would like to know how much of the familiar quasinormal mode/ringdown structure applies to the expulsion of magnetic shear.  What sets the time-scale?

Another reason for doubting CPMS states can persist indefinitely comes from
linearized gravity,  where no realistic sources are known which can generate CPMS behavior for more than finite periods.  Magnetic quadrupoles can be thought of as first moments of the angular momentum density, and this suggests, roughly, that to create $\smag$ (proportional to ${\ddot Q}_{ab}$) one needs to increasingly separate increasingly large contributions to the total angular momentum.  It is possible that such processes can only occur over restricted periods.  This would accord with the general sense of suggestions of Winicour and M\"adler \cite{Winicour2014,MaedlerWinicour2016}, although 
from the present point of view their assumptions seem overly restrictive.

The definition given here of CPMS regimes is idealized in that I have assumed $\dot\sigma =0$.  In a realistic situation, of course, one does not expect this to hold exactly, and one may ask what the consequences of this are.  The tolerance allowed in $\dot\sigma$ will vary according to just what effects are considered, and must be investigated on a case-by-case basis.  This is discussed further in Subsection~\ref{vsscrif}, below.  However, in this paper the exact CPMS condition was only invoked in a couple of places:  as a conceptual issue, in the discussion of magnetic radiation memory; and in Section~VI, the detailed investigation of the far zone of a quadrupole.  It was not important in the induction-zone discussions, or used in the computation of the red-shift effect.

\subsection{Scattering and red-shift}

Two observables associated with the far zones of magnetic quadrupoles in linearized gravity were computed.

One of these was the case of timelike geodesics in a CPMS regime.  We saw there that for a clean far-zone limit the geodesics had to be at least moderately relativistic with respect to the source; that this is at odds with the sorts of configurations often considered for radiation memory.  The polarization effects even in this limit were rather complicated, but the overall magnitude went as $ |\smag |/b$, where $b$ is the impact parameter.  This should be compared to the gravitational scattering due to a mass $\sim M/b$; both of these are effects due to the curvature quantity $\Psi_2$, one from its magnetic part and one from its electric part.

The second case, for which the CPMS condition was not assumed, was for the temporal component of the scattering of light --- the red-shift.  For this, the far-zone magnitude went as $ |\smag (u_0)|/b$, where $u_0$ was the retarded time of receipt of the light.  That the result does not (in the limit of large $b$) depend on $\smag (u)$ for $u\leq u_0$  is a retardation effect; one can view this as an {\em amnesiac} effect, opposite to a memory one.   

This magnetic red-shift would presumably be possible to search for astrophysically; one would look for a central source, and in the circle of directions around the source red-shifts falling off (in the far zone) as $1/b$, with a sinusoidal dependence on the angle around the source.  However, one would also like to know how to rule out other red-shifts of this same form.  I pointed out that if a boosted mass would give something of this form; one could address this by trying to measure the source's velocity.  But it is also possible that higher-multipole electric multipoles could produce this sort of effect; this question needs to be investigated.

\subsection{Magnetic radiation memory}

An important motivation for this paper was the investigation of potential magnetic radiation memory effects.  Previous work of Winicour and M\"adler had tended to suggest these effects could not occur, and simple models of radiation memory had involved electric effects only.  What have we learned?

We have good arguments that in principle magnetic radiation memory effects ought to be possible, at least over finite periods.  Mathematically, they would be due to the transition from one CPMS regime to another.\footnote{Strictly speaking, that is for the case of pure magnetic radiation.  One might also want to allow arbitrary $\sel$.}  However, there are a number of cautionary points:

(a)  We do not have realistic models of sources for significant CPMS behavior, and questions have been raised about how long this behavior can persist. 

(b) Traditional proposals to measure radiation memory are based on observing changes in the trajectories of test-particles from before to after the passage of gravitational waves.  For magnetic effects, because the bracketing CPMS periods are not stationary, the class of trajectories which are cleanly in the far zone is significantly restricted (Section~VI).  In particular, all such trajectories should be relativistic with respect to the source world-line.  This makes the detection of clean magnetic memory effects, even in principle, harder than previous work has suggested.

(c) There is at present little or no prospect of using terrestrial or solar-system gravitational-wave detectors for verifying the existence of magnetic radiation memory.  The issue is that the split of radiation into electric and magnetic parts necessarily involves some comparison of signals at different points of the sphere of asymptotic directions around the source, and we cannot expect current detectors to have the requisite angular resolution.

For instance, the most direct approaches would involve measurements of the radiation field $\sim \Psi _4^0/r$, a spin-weight (minus) two quantity.  And the most local approach to extracting a purely magnetic effect from this would be to compute the spin-weight zero quantity $\Im\eth ^2\Psi _4^0$.  Doing this would require gravitational-wave detectors extending over a large enough solid angle around the source that this second derivative could be accurately found.

(d) On the other hand, there is some prospect of measuring the magnetic red-shift effect around astrophysical sources.  If one can rule out other sources of the red-shift, and if one observed time-independent such red-shifts in two intervals, the difference between them would be a magnetic radiation memory effect.

\subsection{The far zone versus null infinity}\label{vsscrif}

In much of the relativity literature, investigation of the asymptotic regimes is done in the limit of passing to null infinity.  In this paper, however, while some important formulas were derived there, most of the asymptotic work has been done in the far zone.  This was necessary in order to investigate the domains of validity of the computations.

There is, however, a point about the distinction between the far zone and null infinity which has 
not yet been discussed, and which can be puzzling.  One could argue that $\dot\sigma$ is never truly zero, so in particular CPMS regimes cannot truly exist.  If we are indeed considering a Bondi--Sachs space--time, then by passing to large enough $r$ we expect a finite if small radiative term will exist and be the dominant contribution to the curvature.  If this argument were correct, then the CPMS regimes would be a thin set unrepresentative of real physics.

To resolve this, the first observation to make is that the same sort of argument would (for instance) apply to suggest that stationary solutions were unphysical over-idealizations not representative of real physics.  That conclusion would be false, because while we do not expect any real system to be exactly stationary, many systems are adequately modeled by stationary solutions for a wide range of purposes.  It is a question of which aspects of the physics are to be modeled, and how accurate the models must be.

In the case at hand, we should remember that we do not expect any real system to be exactly modeled by a Bondi--Sachs solution.  No system in the Universe is really perfectly isolated.  What we can ask for is that a regime around the system (but not really extending infinitely far out) is well-modeled (for specific purposes) by the leading terms in the Bondi--Sachs expansions, and this has been the definition of the far zone adopted here.  

For any real system, there will be some tolerances in the possible choices of Bondi--Sachs modeling solutions.  In particular, quantities like $\sigma$ and $\Psi _n$ are not precisely determined.  The Bondi--Sachs space--times will be good models if those tolerances are small enough not to affect the analysis of quantities of interest.
So a physical region will be well-modeled by a CPMS regime for certain purposes if the presence of a sufficiently small amount of $\dot\sigma$ does not make any difference to the quantities we wish to model, to the accuracy required.

The point of the discussion just given is only to make precise the {\em sense} in which CPMS regimes might be realistic models of physics.  It does not speak to the dynamical questions of how long the regimes might persist to given degrees of accuracy.

\subsection{Induction-zone effects}

Some consequences of a magnetic quadrupole in the induction zone were investigated.  It was found that memory effects were possible there --- where by memory, we mean differences in the trajectories of particles from ingoing to outgoing regimes which signal that at some point the quadrupole was non-zero.  That such effects are possible is not a surprise; but we do find some chance that they could be verified by laboratory experiments.  They could also lead to astrophysical effects, but these seem less likely to be observed, at least based on current understanding and technology.

Arguably the most conceptually interesting result was that a magnetic gravitational quadrupole will tend to induce, in nearby test particles, motions leading to an {\em opposing} quadrupole.  
This is perhaps the first example of how matter might tend to screen magnetic general-relativistic effects.  

While this effect has the general flavor of Lenz's law for electromagnetism, the parallel is not very close.  Lenz's law describes currents induced by a changing field, whereas here the test particles react to the value of the quadrupole (not some time-derivative of that).

\begin{acknowledgments}

I thank Jeffrey Winicour for stimulating conversations about magnetic shear.  I know he was aware of the two-hoop example a number of years ago.

\end{acknowledgments}
  
\section*{Appendix:  Scattering Integrals}

I will here indicate how the integrals occurring in the scattering computations can be done.

Each integral can be reduced to a sum of ones of the form
\begin{eqnarray}
  I= \int _{-\infty}^\infty \frac{s^m}{(s^2+b^2)^{n/2}} q(t_0 +s-\sqrt{s^2+b^2})\, ds\, ,
\end{eqnarray}
where $s$ is an affine parameter on the geodesic and $q$ is a component of the quadrupole, or of a derivative of the quadrupole, and $n-m\geq 2$.

The key point is the assumption that $q(u)$ is non-zero only for a finite range of retarded times $u$.  In fact, if we change variables to $u=t_0+s-\sqrt{s^2+b^2}$, we find that $u$ ranges over the interval $(-\infty ,t_0]$.  We invert the relation to get
\begin{eqnarray}
  s &=&\frac{b^2}{2(t_0-u)} -\frac{t_0-u}{2}\, .
\end{eqnarray}
Note here that, if $u$ is restricted to any bounded subinterval of $(-\infty ,t_0]$, then 
\begin{eqnarray}
  \frac{s^m}{(s^2+b^2)^{n/2}}
  &=& \left( \frac{2(t_0-u)}{b^2} \right)^{n-m}
    +\cdots
\end{eqnarray}
as $b\to\infty$, uniformly for in $u$ in any bounded interval (for $n-m\geq 0$).  

We will also have
\begin{eqnarray}
  ds =\left( \frac{b^2}{2(t_0-u)^2} +\frac{1}{2}\right)\, du\, .
\end{eqnarray}  
Then 
\begin{eqnarray}
I&=& (b^2/2)^{(m-n+1)} \int _{-\infty}^{u_0} (t_0-u)^{n-m-2} q(u)\, du +\cdots\, .\qquad
\end{eqnarray}



\bibliography{qmtprd.revc.bbl}

\begin{thebibliography}{23}%
\makeatletter
\providecommand \@ifxundefined [1]{%
 \@ifx{#1\undefined}
}%
\providecommand \@ifnum [1]{%
 \ifnum #1\expandafter \@firstoftwo
 \else \expandafter \@secondoftwo
 \fi
}%
\providecommand \@ifx [1]{%
 \ifx #1\expandafter \@firstoftwo
 \else \expandafter \@secondoftwo
 \fi
}%
\providecommand \natexlab [1]{#1}%
\providecommand \enquote  [1]{``#1''}%
\providecommand \bibnamefont  [1]{#1}%
\providecommand \bibfnamefont [1]{#1}%
\providecommand \citenamefont [1]{#1}%
\providecommand \href@noop [0]{\@secondoftwo}%
\providecommand \href [0]{\begingroup \@sanitize@url \@href}%
\providecommand \@href[1]{\@@startlink{#1}\@@href}%
\providecommand \@@href[1]{\endgroup#1\@@endlink}%
\providecommand \@sanitize@url [0]{\catcode `\\12\catcode `\$12\catcode
  `\&12\catcode `\#12\catcode `\^12\catcode `\_12\catcode `\%12\relax}%
\providecommand \@@startlink[1]{}%
\providecommand \@@endlink[0]{}%
\providecommand \url  [0]{\begingroup\@sanitize@url \@url }%
\providecommand \@url [1]{\endgroup\@href {#1}{\urlprefix }}%
\providecommand \urlprefix  [0]{URL }%
\providecommand \Eprint [0]{\href }%
\providecommand \doibase [0]{http://dx.doi.org/}%
\providecommand \selectlanguage [0]{\@gobble}%
\providecommand \bibinfo  [0]{\@secondoftwo}%
\providecommand \bibfield  [0]{\@secondoftwo}%
\providecommand \translation [1]{[#1]}%
\providecommand \BibitemOpen [0]{}%
\providecommand \bibitemStop [0]{}%
\providecommand \bibitemNoStop [0]{.\EOS\space}%
\providecommand \EOS [0]{\spacefactor3000\relax}%
\providecommand \BibitemShut  [1]{\csname bibitem#1\endcsname}%
\let\auto@bib@innerbib\@empty
\bibitem [{\citenamefont {Helfer}(2007)}]{ADH2007}%
  \BibitemOpen
  \bibfield  {author} {\bibinfo {author} {\bibfnamefont {A.~D.}\ \bibnamefont
  {Helfer}},\ }\href@noop {} {\bibfield  {journal} {\bibinfo  {journal} {Gen.
  Rel. Grav.}\ }\textbf {\bibinfo {volume} {39}},\ \bibinfo {pages} {2125}
  (\bibinfo {year} {2007})}\BibitemShut {NoStop}%
\bibitem [{\citenamefont {{Teukolsky}}(1972)}]{Teukolsky1972}%
  \BibitemOpen
  \bibfield  {author} {\bibinfo {author} {\bibfnamefont {S.~A.}\ \bibnamefont
  {{Teukolsky}}},\ }\href {\doibase 10.1103/PhysRevLett.29.1114} {\bibfield
  {journal} {\bibinfo  {journal} {Physical Review Letters}\ }\textbf {\bibinfo
  {volume} {29}},\ \bibinfo {pages} {1114} (\bibinfo {year}
  {1972})}\BibitemShut {NoStop}%
\bibitem [{\citenamefont {Bieri}\ and\ \citenamefont
  {Garfinkle}(2014)}]{BieriGarfinkle2014}%
  \BibitemOpen
  \bibfield  {author} {\bibinfo {author} {\bibfnamefont {L.}~\bibnamefont
  {Bieri}}\ and\ \bibinfo {author} {\bibfnamefont {D.}~\bibnamefont
  {Garfinkle}},\ }\href {\doibase 10.1103/PhysRevD.89.084039} {\bibfield
  {journal} {\bibinfo  {journal} {Phys. Rev. D}\ }\textbf {\bibinfo {volume}
  {89}},\ \bibinfo {pages} {084039} (\bibinfo {year} {2014})}\BibitemShut
  {NoStop}%
\bibitem [{\citenamefont {M{\"a}dler}\ and\ \citenamefont
  {Winicour}(2016)}]{MaedlerWinicour2016}%
  \BibitemOpen
  \bibfield  {author} {\bibinfo {author} {\bibfnamefont {T.}~\bibnamefont
  {M{\"a}dler}}\ and\ \bibinfo {author} {\bibfnamefont {J.}~\bibnamefont
  {Winicour}},\ }\href {\doibase 10.1088/0264-9381/33/17/175006} {\bibfield
  {journal} {\bibinfo  {journal} {Classical and Quantum Gravity}\ }\textbf
  {\bibinfo {volume} {33}},\ \bibinfo {pages} {175006} (\bibinfo {year}
  {2016})}\BibitemShut {NoStop}%
\bibitem [{\citenamefont {Satishchandran}\ and\ \citenamefont
  {Wald}(2019)}]{SW2019a}%
  \BibitemOpen
  \bibfield  {author} {\bibinfo {author} {\bibfnamefont {G.}~\bibnamefont
  {Satishchandran}}\ and\ \bibinfo {author} {\bibfnamefont {R.~M.}\
  \bibnamefont {Wald}},\ }\href {\doibase 10.1103/PhysRevD.99.084007}
  {\bibfield  {journal} {\bibinfo  {journal} {Phys. Rev. D}\ }\textbf {\bibinfo
  {volume} {99}},\ \bibinfo {pages} {084007} (\bibinfo {year}
  {2019})}\BibitemShut {NoStop}%
\bibitem [{\citenamefont {{Akiyama}}\ and\ \citenamefont {others {(Event
  Horizon Telescope Collaboration)}}(2019)}]{EHT1ab}%
  \BibitemOpen
  \bibfield  {author} {\bibinfo {author} {\bibfnamefont {K.}~\bibnamefont
  {{Akiyama}}}\ and\ \bibinfo {author} {\bibnamefont {others {(Event Horizon
  Telescope Collaboration)}}},\ }\href {\doibase 10.3847/2041-8213/ab0ec7}
  {\bibfield  {journal} {\bibinfo  {journal} {\apj}\ }\textbf {\bibinfo
  {volume} {875}},\ \bibinfo {eid} {L1} (\bibinfo {year} {2019})}\BibitemShut
  {NoStop}%
\bibitem [{\citenamefont {Johannsen}\ and\ \citenamefont
  {Psaltis}(2010)}]{JP2010}%
  \BibitemOpen
  \bibfield  {author} {\bibinfo {author} {\bibfnamefont {T.}~\bibnamefont
  {Johannsen}}\ and\ \bibinfo {author} {\bibfnamefont {D.}~\bibnamefont
  {Psaltis}},\ }\href {\doibase 10.1088/0004-637x/718/1/446} {\bibfield
  {journal} {\bibinfo  {journal} {The Astrophysical Journal}\ }\textbf
  {\bibinfo {volume} {718}},\ \bibinfo {pages} {446} (\bibinfo {year}
  {2010})}\BibitemShut {NoStop}%
\bibitem [{\citenamefont {Penrose}\ and\ \citenamefont
  {Rindler}(1984)}]{PR1984}%
  \BibitemOpen
  \bibfield  {author} {\bibinfo {author} {\bibfnamefont {R.}~\bibnamefont
  {Penrose}}\ and\ \bibinfo {author} {\bibfnamefont {W.}~\bibnamefont
  {Rindler}},\ }\href@noop {} {\emph {\bibinfo {title} {Spinors and
  space--time, vol. 1: Two--spinor calculus and relativistic fields}}}\
  (\bibinfo  {publisher} {Cambridge University Press},\ \bibinfo {year}
  {1984})\BibitemShut {NoStop}%
\bibitem [{\citenamefont {Penrose}\ and\ \citenamefont
  {Rindler}(1986)}]{PR1986}%
  \BibitemOpen
  \bibfield  {author} {\bibinfo {author} {\bibfnamefont {R.}~\bibnamefont
  {Penrose}}\ and\ \bibinfo {author} {\bibfnamefont {W.}~\bibnamefont
  {Rindler}},\ }\href@noop {} {\emph {\bibinfo {title} {Spinors and
  space--time, vol. 2: Spinor and twistor methods in space--time geometry}}}\
  (\bibinfo  {publisher} {Cambridge University Press},\ \bibinfo {year}
  {1986})\BibitemShut {NoStop}%
\bibitem [{\citenamefont {Newman}\ and\ \citenamefont
  {Penrose}(1966)}]{NP1966}%
  \BibitemOpen
  \bibfield  {author} {\bibinfo {author} {\bibfnamefont {E.~T.}\ \bibnamefont
  {Newman}}\ and\ \bibinfo {author} {\bibfnamefont {R.}~\bibnamefont
  {Penrose}},\ }\href {\doibase 10.1063/1.1931221} {\bibfield  {journal}
  {\bibinfo  {journal} {Journal of Mathematical Physics}\ }\textbf {\bibinfo
  {volume} {7}},\ \bibinfo {pages} {863} (\bibinfo {year} {1966})},\ \Eprint
  {http://arxiv.org/abs/https://doi.org/10.1063/1.1931221}
  {https://doi.org/10.1063/1.1931221} \BibitemShut {NoStop}%
\bibitem [{\citenamefont {Bondi}\ \emph {et~al.}(1962)\citenamefont {Bondi},
  \citenamefont {van~der Burg},\ and\ \citenamefont {Metzner}}]{BVM}%
  \BibitemOpen
  \bibfield  {author} {\bibinfo {author} {\bibfnamefont {H.}~\bibnamefont
  {Bondi}}, \bibinfo {author} {\bibfnamefont {M.~G.~J.}\ \bibnamefont {van~der
  Burg}}, \ and\ \bibinfo {author} {\bibfnamefont {A.~W.~K.}\ \bibnamefont
  {Metzner}},\ }\href@noop {} {\bibfield  {journal} {\bibinfo  {journal} {Proc.
  R. Soc. Lond.}\ }\textbf {\bibinfo {volume} {A269}},\ \bibinfo {pages} {21}
  (\bibinfo {year} {1962})}\BibitemShut {NoStop}%
\bibitem [{\citenamefont {Penrose}(2011)}]{Penrose1964}%
  \BibitemOpen
  \bibfield  {author} {\bibinfo {author} {\bibfnamefont {R.}~\bibnamefont
  {Penrose}},\ }\bibfield  {booktitle} {\emph {\bibinfo {booktitle}
  {{Relativit{\'e}, Groupes et Topologie: Proceedings, Ecole d'{\'e}t{\'e} de
  Physique Th{\'e}orique, Session XIII, Les Houches, France, Jul 1 - Aug 24,
  1963}}},\ }\href {\doibase 10.1007/s10714-010-1110-5} {\bibfield  {journal}
  {\bibinfo  {journal} {Gen. Rel. Grav.}\ }\textbf {\bibinfo {volume} {43}},\
  \bibinfo {pages} {901} (\bibinfo {year} {2011})},\ \bibinfo {note}
  {[,565(1964)]}\BibitemShut {NoStop}%
\bibitem [{\citenamefont {Newman}\ and\ \citenamefont
  {Penrose}(1962)}]{NP1961}%
  \BibitemOpen
  \bibfield  {author} {\bibinfo {author} {\bibfnamefont {E.}~\bibnamefont
  {Newman}}\ and\ \bibinfo {author} {\bibfnamefont {R.}~\bibnamefont
  {Penrose}},\ }\href {\doibase 10.1063/1.1724257} {\bibfield  {journal}
  {\bibinfo  {journal} {J.Math.Phys.}\ }\textbf {\bibinfo {volume} {3}},\
  \bibinfo {pages} {566} (\bibinfo {year} {1962})}\BibitemShut {NoStop}%
\bibitem [{\citenamefont {{Zel'dovich}}\ and\ \citenamefont
  {{Polnarev}}(1974)}]{ZeldovichPolnarev1974}%
  \BibitemOpen
  \bibfield  {author} {\bibinfo {author} {\bibfnamefont {Y.~B.}\ \bibnamefont
  {{Zel'dovich}}}\ and\ \bibinfo {author} {\bibfnamefont {A.~G.}\ \bibnamefont
  {{Polnarev}}},\ }\href@noop {} {\bibfield  {journal} {\bibinfo  {journal}
  {{Sov. Astron.}}\ }\textbf {\bibinfo {volume} {18}},\ \bibinfo {pages} {17}
  (\bibinfo {year} {1974})}\BibitemShut {NoStop}%
\bibitem [{\citenamefont {Braginsky}\ and\ \citenamefont
  {Grishchuk}(1985)}]{BraginskyGrishchuk1985}%
  \BibitemOpen
  \bibfield  {author} {\bibinfo {author} {\bibfnamefont {V.~B.}\ \bibnamefont
  {Braginsky}}\ and\ \bibinfo {author} {\bibfnamefont {L.~P.}\ \bibnamefont
  {Grishchuk}},\ }\href@noop {} {\bibfield  {journal} {\bibinfo  {journal}
  {Sov. Phys. JETP}\ }\textbf {\bibinfo {volume} {62}},\ \bibinfo {pages} {427}
  (\bibinfo {year} {1985})},\ \bibinfo {note} {[Zh. Eksp. Teor.
  Fiz.89,744(1985)]}\BibitemShut {NoStop}%
\bibitem [{\citenamefont {{Braginskii}}\ and\ \citenamefont
  {{Thorne}}(1987)}]{BraginskyThorne1987}%
  \BibitemOpen
  \bibfield  {author} {\bibinfo {author} {\bibfnamefont {V.~B.}\ \bibnamefont
  {{Braginskii}}}\ and\ \bibinfo {author} {\bibfnamefont {K.~S.}\ \bibnamefont
  {{Thorne}}},\ }\href {\doibase 10.1038/327123a0} {\bibfield  {journal}
  {\bibinfo  {journal} {\nat}\ }\textbf {\bibinfo {volume} {327}},\ \bibinfo
  {pages} {123} (\bibinfo {year} {1987})}\BibitemShut {NoStop}%
\bibitem [{\citenamefont {{Christodoulou}}(1991)}]{Christodoulou1991}%
  \BibitemOpen
  \bibfield  {author} {\bibinfo {author} {\bibfnamefont {D.}~\bibnamefont
  {{Christodoulou}}},\ }\href {\doibase 10.1103/PhysRevLett.67.1486} {\bibfield
   {journal} {\bibinfo  {journal} {Physical Review Letters}\ }\textbf {\bibinfo
  {volume} {67}},\ \bibinfo {pages} {1486} (\bibinfo {year}
  {1991})}\BibitemShut {NoStop}%
\bibitem [{\citenamefont {Helfer}(2016)}]{ADH2016FGG}%
  \BibitemOpen
  \bibfield  {author} {\bibinfo {author} {\bibfnamefont {A.~D.}\ \bibnamefont
  {Helfer}},\ }\href {\doibase 10.1103/PhysRevD.94.124011} {\bibfield
  {journal} {\bibinfo  {journal} {Phys. Rev. D}\ }\textbf {\bibinfo {volume}
  {94}},\ \bibinfo {pages} {124011} (\bibinfo {year} {2016})}\BibitemShut
  {NoStop}%
\bibitem [{\citenamefont {{Helfer}}(2018)}]{ADH2018}%
  \BibitemOpen
  \bibfield  {author} {\bibinfo {author} {\bibfnamefont {A.~D.}\ \bibnamefont
  {{Helfer}}},\ }\href@noop {} {\bibfield  {journal} {\bibinfo  {journal}
  {arXiv e-prints}\ } (\bibinfo {year} {2018})},\ \Eprint
  {http://arxiv.org/abs/1805.11569} {arXiv:1805.11569 [gr-qc]} \BibitemShut
  {NoStop}%
\bibitem [{\citenamefont {{Helfer}}(2013)}]{ADH2013}%
  \BibitemOpen
  \bibfield  {author} {\bibinfo {author} {\bibfnamefont {A.~D.}\ \bibnamefont
  {{Helfer}}},\ }\href {\doibase 10.1093/mnras/sts618} {\bibfield  {journal}
  {\bibinfo  {journal} {Mon. Not. R. Astron. Soc.}\ }\textbf {\bibinfo {volume}
  {430}},\ \bibinfo {pages} {305} (\bibinfo {year} {2013})},\ \Eprint
  {http://arxiv.org/abs/1212.2926} {arXiv:1212.2926 [gr-qc]} \BibitemShut
  {NoStop}%
\bibitem [{\citenamefont {{Helfer}}(2014)}]{ADH2014}%
  \BibitemOpen
  \bibfield  {author} {\bibinfo {author} {\bibfnamefont {A.~D.}\ \bibnamefont
  {{Helfer}}},\ }\href {\doibase 10.1103/PhysRevD.90.044005} {\bibfield
  {journal} {\bibinfo  {journal} {\prd}\ }\textbf {\bibinfo {volume} {90}},\
  \bibinfo {eid} {044005} (\bibinfo {year} {2014})},\ \Eprint
  {http://arxiv.org/abs/1407.2816} {arXiv:1407.2816 [gr-qc]} \BibitemShut
  {NoStop}%
\bibitem [{\citenamefont {Lasky}\ \emph {et~al.}(2016)\citenamefont {Lasky},
  \citenamefont {Thrane}, \citenamefont {Levin}, \citenamefont {Blackman},\
  and\ \citenamefont {Chen}}]{LTLB2016}%
  \BibitemOpen
  \bibfield  {author} {\bibinfo {author} {\bibfnamefont {P.~D.}\ \bibnamefont
  {Lasky}}, \bibinfo {author} {\bibfnamefont {E.}~\bibnamefont {Thrane}},
  \bibinfo {author} {\bibfnamefont {Y.}~\bibnamefont {Levin}}, \bibinfo
  {author} {\bibfnamefont {J.}~\bibnamefont {Blackman}}, \ and\ \bibinfo
  {author} {\bibfnamefont {Y.}~\bibnamefont {Chen}},\ }\href {\doibase
  10.1103/PhysRevLett.117.061102} {\bibfield  {journal} {\bibinfo  {journal}
  {Phys. Rev. Lett.}\ }\textbf {\bibinfo {volume} {117}},\ \bibinfo {pages}
  {061102} (\bibinfo {year} {2016})}\BibitemShut {NoStop}%
\bibitem [{\citenamefont {{Winicour}}(2014)}]{Winicour2014}%
  \BibitemOpen
  \bibfield  {author} {\bibinfo {author} {\bibfnamefont {J.}~\bibnamefont
  {{Winicour}}},\ }\href {\doibase 10.1088/0264-9381/31/20/205003} {\bibfield
  {journal} {\bibinfo  {journal} {Classical and Quantum Gravity}\ }\textbf
  {\bibinfo {volume} {31}},\ \bibinfo {eid} {205003} (\bibinfo {year}
  {2014})},\ \Eprint {http://arxiv.org/abs/1407.0259} {arXiv:1407.0259 [gr-qc]}
  \BibitemShut {NoStop}%
\end{thebibliography}%

\end{document}